\documentclass{article}
\usepackage{authblk}
\usepackage[linktocpage]{hyperref}
\usepackage{amsmath}
\usepackage{amsthm}
\usepackage{amssymb}
\usepackage{mathtools}
\usepackage{bbold}
\usepackage{graphicx}
\usepackage{float}
\usepackage{wrapfig}
\usepackage{color}
\usepackage{dsfont}
\usepackage{float}
\usepackage{floatflt}
\usepackage{hhline}
\usepackage{hyperref}
\usepackage{amsfonts}
\usepackage{comment}
\usepackage{braket}
\usepackage{csquotes}
\usepackage{accents}
\usepackage{enumitem}
\usepackage{cancel}
\usepackage{subcaption}
\usepackage{enumitem}
\usepackage{cleveref}
\usepackage[normalem]{ulem}
\usepackage{newunicodechar}
\newunicodechar{α}{\ensuremath{\alpha}}
 \definecolor{darkblue}{RGB}{0,0,149}
\hypersetup{colorlinks=true,linkcolor=darkblue,filecolor=darkblue,urlcolor=darkblue,citecolor=darkblue,linktocpage=true}
\usepackage{url}
\usepackage{lineno}
\newtheorem{definition}{Definition}

\newtheorem{statement}{Statement}

\DeclareMathOperator{\Tr}{Tr}

\title{The classical limit of quantum mechanics through coarse-grained measurements}

\author[1,2,3]{\small Fatemeh Bibak\thanks{These authors contributed equally to this work.}}

\author[*1,2,3]{Carlo Cepollaro}
\author[1,3]{Nicol\'as Medina S\'anchez}
\author[1,2]{Borivoje Daki\'c}
\author[1,2]{\v Caslav Brukner}

\affil[1]{\small University of Vienna, Faculty of Physics, Vienna Center for Quantum Science and Technology, Boltzmanngasse 5, 1090 Vienna, Austria}
\affil[2]{\small Institute for Quantum Optics and Quantum Information (IQOQI), Austrian Academy of Sciences, Boltzmanngasse 3, A-1090 Vienna, Austria}
\affil[3]{\small University of Vienna, Vienna Doctoral School in Physics, Boltzmanngasse 5, 1090 Vienna, Austria}

\begin{document}

\date{\vspace{-5ex}}
\maketitle
\begin{abstract}

Understanding how classical physics emerges from quantum mechanics remains a central problem in the foundations of physics. Here we derive a classical limit from finite-resolution measurements, modeled by continuous coarse-grained POVMs. When the resolved phase-space area is large compared with Planck’s constant, the accessible statistics of any quantum state admit an effective classical description: coarse-grained observables become approximately jointly measurable and the induced probability density is positive. We derive the exact evolution equation for this density and show that, in the strong coarse-graining regime, its non-Liouville corrections are suppressed up to an Ehrenfest time, resulting in classical Hamiltonian flow generated by a Hamiltonian smoothed over the measurement cell. When the Hamiltonian varies negligibly across such a cell, the smoothed Hamiltonian reduces to the classical Hamiltonian whose quantization produced the quantum dynamics, thereby closing the quantization--classical-limit loop. Repeated finite-resolution measurements then generate stochastic records confined, with high probability, to tubes around classical trajectories. Our results provide a unified operational framework for the quantum-to-classical transition in microscopic and macroscopic systems, and establish the consistency of the quantization--classical-limit cycle.

\end{abstract}
\newpage
\tableofcontents

\section{Introduction}\label{sec:Introduction}

Our everyday experience provides clear evidence for the validity of the classical physical description at both the kinematic and dynamic levels. Does this description emerge from a more fundamental theory? And, if so, how? These questions are still not fully resolved in physics, despite considerable efforts and a vast literature on the classical limit \cite{ehrenfest1927,Peres1995,landsman2005, layton_oppenheim2024}. It is sometimes believed that the problem can be solved by formal derivations in which the limit of Planck’s constant is taken to be zero, i.e. $\hbar \rightarrow 0$. But this cannot be an answer, because $\hbar$ is a dimensional physical constant different from zero, and yet classical physics is an accurate description in the realm of everyday life. Any operationally meaningful approach to the classical limit must therefore keep $\hbar$ fixed.

There are two main approaches to understanding the origin of classical physics. In the first approach, it is generally assumed that the classical limit cannot be derived without modifying quantum theory. An example is the introduction of additional collapse mechanisms of quantum states \cite{bassi_collapse_2023}, assumed to be due either to a large number of degrees of freedom in a macroscopic superposition
\cite{ghirardi_unified_1986}, or due to the supposed gravitational self-interaction \cite{diosi_universal_1987,penrose_gravitys_1996}. In the second approach, one derives classical physics solely as a limit of quantum theory. Classical physics then arises either through decoherence \cite{zeh_interpretation_1970,zurek_pointer_1981,zurek1982,joos1985,zurek2003}, i.e. interaction between the system with its environment, or through coarse-grained observation of the system \cite{omnes_logical_1989,Peres1995,kofler2007}. 
The decoherence approach provides a dynamical perspective on the quantum-to-classical transition, emphasizing how interactions with the environment lead to the loss of quantum coherence in a system, thereby inducing classical behavior. Building on this, the framework of Quantum Darwinism~\cite{Zurek2009QuantumDarwinism,Zurek2007EnvironmentWitness,BlumeKohoutZurek2015ObjectiveReality} further explains how classical objectivity of states and records emerges from the redundant encoding of information about the system into many environmental degrees of freedom.
Complementary to the decoherence program, the transition to classicality via coarse-grained measurements highlights an epistemic perspective, focusing on what can and cannot be observed from the system under limited measurement precision, even if the system's quantum state is fully coherent. 

A particularly instructive operational probe of the quantum-to-classical transition is provided by matter-wave interferometry with increasingly massive molecules and nanoparticles, as demonstrated in Refs.~\cite{Pedalino2025NanoparticleCats,Eibenberger2013Over10000amu}. These experiments can be seen as a testbed for various approaches to the classical limit. Firstly, they demonstrate, for a wide range of masses, agreement with quantum-mechanical predictions (although the parameter regime where collapse models make distinct predictions has not yet been reached)~\cite{bassi_collapse_2023,Vinante2017CantileverCSL,Helou2017LISAPathfinderCSL,Pedalino2025NanoparticleCats}. Secondly, they also show loss of interference visibility and a transition towards classicality due to decoherence, either through radiation by heating the molecules on their way through the interferometer~\cite{Hackermueller2004ThermalEmissionDecoherence} or through collisions with residual gas molecules in the vacuum chamber~\cite{Hornberger2003CollisionalDecoherence}. Finally, they show that even with negligible decoherence the measured statistics can become compatible with classical trajectories through the slits and subsequent detection on the screen as the molecular mass increases, if the spatial resolution of the detector is no longer sufficient to resolve the increasingly fine interference fringes~\cite{Pedalino2025NanoparticleCats}.

We thus see that ``classical behaviour'' can arise over a broad range of experimental parameters, either because continuous monitoring induces substantial decoherence, or because the available measurement precision is insufficient to resolve genuinely quantum features. Despite their differing emphases, the decoherence and coarse-grained approaches are deeply interconnected. First, a purification of a coarse-grained measurement, via Naimark’s theorem, can always be interpreted as an interaction between the system and additional degrees of freedom of the measuring device, which effectively act as an environment, followed by a projective measurement on these environmental degrees of freedom. Conversely, every operational statement in quantum mechanics must inherently involve not only a quantum state, but also a measurement to provide  probabilities via Born's rule. This applies also to the states arising typically in the interaction with environment: as long as the joint system-environment system undergoes a unitary quantum mechanical evolution, there exists, in principle, a measurement on the joint state that could reveal its quantum nature, even if such a measurement is completely unrealistic in practice. That realistic measurements of the environment typically correspond to coarse-grained observations of the system is precisely what makes the two approaches complementary rather than competing. In this paper, we explore coarse-grained measurements as a framework for understanding the quantum-to-classical transition, acknowledging their interconnection with the decoherence program, yet emphasizing the interplay between the measurement device's phase space precision and the system's inherent quantum uncertainty within phase space.

One of the most pivotal tools for evaluating the compatibility of quantum mechanical systems with classical physics is the set of Bell~\cite{bell_epr_1964} and Leggett-Garg (LG)~\cite{leggett1985,leggett2002} inequalities. The Bell inequalities impose constraints
on spatial correlations at a given time, ensuring consistency with the principle
of local causality~\cite{brunner_bell_2014}, which serves as a formalization of the kinematics
inherent in classical systems. Analogously, the LG inequalities are constraints on temporal correlations that are derived from a conjunction of two assumptions, \textit{macrorealism per se} and \textit{non-invasive measurability}. Macrorealism per se is the assumption that a macroscopic physical quantity, for example the position of a macroscopic object, has a definite value at any given time prior to and independently of the measurement. Non-invasive measurability is the assumption that the disturbance caused by the measurement of the macroscopic quantity can be assumed to be arbitrarily small. A prime example of a theory that respects both assumptions is classical mechanics and classical statistical mechanics, which evolve according to Liouville dynamics in phase space or through classical stochastic processes. Quantum mechanical correlations are known to violate both Bell and LG inequalities~\cite{vitagliano_leggett-garg_2023, Clemente2015}; however, experimental demonstrations of these violations have so far been limited to the microscopic system~\cite{Knee2012,Zhan2023,Dressel2011,Goggin2011,Palacios-Laloy2010,vitelli_quantum_2010}.

Although significant progress has been made in understanding the origins of classical physics through coarse-grained measurements \cite{omnes_logical_1989,GellMann1993,Peres1995,kofler2007}, many questions remain unanswered. 
In coarse-graining approaches, measurement results are grouped into bins, where each bin represents a single measurement outcome obtained by grouping together a collection of \textit{microscopic} outcomes into one effective \textit{macroscopic} result. While it has been shown that coarse-graining outcomes into bins can account for classicality at the kinematical level in specific systems (i.e., a joint probability distribution exists at a given time for all coarse-grained measurements), it is recognized that this explanation depends critically on the specific type of coarse-graining and the measurement precision used -- particularly, the way in which outcomes are grouped into bins. To be precise, compatibility with classical kinematics can only be achieved when measurement outcomes that are adjacent to each other are grouped together in bins \cite{omnes_logical_1989,kofler2007,raeisi_coarse_2011,jeong_coarsening_2014, sekatski_how_2014,Kabernik2021}. In contrast, when non-adjacent outcomes are grouped together, such as in parity measurements in quantum optics, where every odd and every even outcome in the photon-number (Fock) states are grouped separately, they can be exploited to violate the LG~\cite{kofler2007} or Bell’s inequalities even if Wigner’s quasiprobability in phase space is positive~\cite{banaszek_nonlocality_1998}.

Regarding measurement precision, it has been shown that for macroscopic systems composed of many particles, a certain level of precision allows for a classical description of measurement outcomes~\cite{Gallego2021}. This phenomenon is closely related to macroscopic locality~\cite{Navascues2010}: at a specific coarse-graining threshold, the measurement statistics of certain macroscopic states—specifically, those composed of many independent and identically distributed (IID) correlated pairs—exhibit local correlations, meaning they do not violate Bell inequalities. However, further research~\cite{Gallego2021,Gallego2022,Gallego2024} has demonstrated that nonlocal correlations can emerge when the IID assumption is lifted.

Additionally, classicality has been shown to emerge in the mathematical limit of an infinite number of quantum systems, which can be described using a non-separable Hilbert space, i.e. one with uncountably infinite dimensions. In this framework, the Hilbert space decomposes into superselection sectors that cannot interfere with each other, enforcing a classical-like behavior~\cite{VanDenBossche2023a,VanDenBossche2023b,VanDenBossche2023c}.

Moving now to dynamics, we note that even under coarse-graining with bins containing adjacent outcomes, a violation of the LG inequalities and a departure from classical dynamics become possible under ``non-classical'' time evolutions. These evolutions are typically driven by Hamiltonians that generate superpositions between different bins \cite{kofler_conditions_2008, jeong_failure_2009}. Specifically, if we consider coarse-grained measurements in the phase space, all Hamiltonians -- except that of a harmonic oscillator -- will produce superpositions across multiple bins after sufficient time. This raises the question: how can general Liouville dynamics of the classical physics emerge from the underlying quantum mechanical evolution under the coarse-grained measurements?

\begin{figure}[htb!]
\centering
\includegraphics[width=0.55\linewidth]{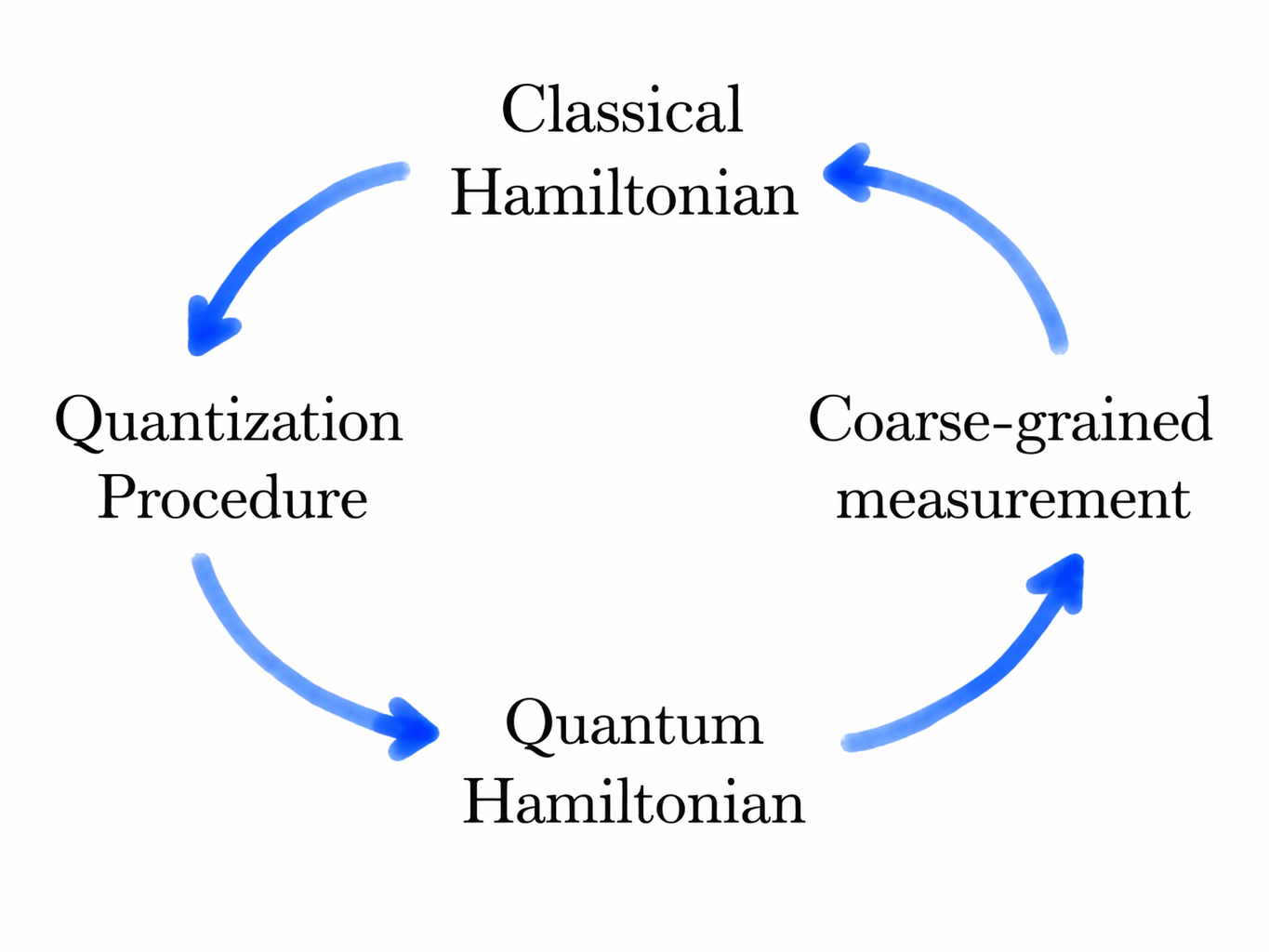}
\caption{The loop consisting of two paths connecting classical and quantum physics in opposite directions. From classical to quantum physics, quantum Hamiltonian is derived from the corresponding classical Hamiltonian using the Dirac quantization procedure. Conversely, in the transition from quantum to classical physics through coarse-grained observation, the effective evolution becomes classical, with an associated effective classical Hamiltonian. A natural question arises: Is the loop consistent? Specifically, is the Hamiltonian that arises from the classical limit the same classical Hamiltonian that is the starting point of the quantization procedure?}
\label{fig:1}
\end{figure}

A unified and comprehensive account of the quantum-to-classical transition—encompassing both the emergence of classical states and their effective classical dynamics across all the classes of situations—has yet to be achieved. To move towards such an understanding, the approach to the classical limit must address the following questions:

\begin{enumerate}
    \item Every coarse-grained observation is characterized by its precision (the size of the measurement bin) and the rate of the measurements. \textit{Is there a range of these parameters under which the effective evolution of the quantum-mechanical system can be described by a classical Hamiltonian?} If so, what form would this Hamiltonian take in the classical limit? Is it possible to generate all possible classical Hamiltonian systems in this way? 

    \item Quantum Hamiltonians are obtained from classical Hamiltonians by Dirac's quantisation rule. {\it Is the effective classical Hamiltonian, which is obtained in the classical limit, the same as the one from which the quantum Hamiltonian was generated by the quantisation procedure?} In other words, is the circle of quantisation-classical limit-quantisation from Figure~\ref{fig:1} consistent?
    \item 
    One distinguishes between two types of classical systems: (a) macroscopic systems, such as chairs, and cats, which appear to inherently exhibit classical behavior under standard physical conditions; and (b) microscopic systems, such as atoms and electrons, which display classical behavior only under specific conditions. A typical example of the latter are particles leaving traces in the form of classical trajectories in a cloud chamber. \textit{Can the classical limit achieved through coarse-grained measurements account for both types of classicality?}
    
\end{enumerate}

Providing comprehensive answers to questions 1–3 would be highly significant for both foundational insights and practical applications in physics. 
From a foundational perspective, this research can provide deeper insight into how classical Newtonian mechanics emerges from quantum mechanics or, alternatively, help identify the essential components of one theory required to formulate the other, thereby ensuring the consistency of the cycle depicted in Figure~\ref{fig:1}. In that respect, it is relevant to mention that while Bohr~\cite{Bohr1949} and the Copenhagen school argued for the necessity of classical physics in explaining quantum mechanics -- asserting that classical concepts provide the necessary language for interpreting measurements -- Bell~\cite{Bell1987} and others contended that quantum theory should be self-sufficient, with any classical behavior emerging solely from an underlying quantum framework. Deriving the full cycle of Fig.~\ref{fig:1} would show that both views are, in fact, compatible: the ``Bohr-like'' and ``Bell-like'' standpoints correspond to different choices of where one prefers to cut and declare the ``beginning'' and the ``end'' of a single closed classical--quantum--classical explanatory loop.

From an applied perspective, understanding dynamics that do not yield a classical limit, even with limited measurement accuracy, is critical to designing experiments that probe macroscopic quantum phenomena \cite{leggett1985,Arndt1999,Friedman2000,OConnell2010,Gerlich2011,Delic2020}. Specifically, understanding the quantum to classical transition can have important practical
applications in quantum technology, in particular for the development of
hybrid quantum--classical systems where a macroscopic object interacts
with a microscopic one~\cite{PeresTerno2001HybridDynamics,Oppenheim2002InconsistencyHybrid,OppenheimSparaciariSodaWellerDavies2022TwoClasses}. Additionally, non-classical dynamics serve as a valuable resource in quantum information
science: it is known that quantum dynamics are hard to simulate with classical systems, and
that they can save resources in memory size in temporal analogues of communication-complexity
protocols~\cite{brukner2004,kleinmann_cost_2011}.

\section{The State-of-Art}

The previous works have addressed some of the questions given above while leaving others aside, or have provided partial answers. As discussed in the Introduction, decoherence~\cite{zeh_interpretation_1970,zurek_pointer_1981,zurek1982,joos1985,zurek2003} and Quantum Darwinism~\cite{Zurek2009QuantumDarwinism,Zurek2007EnvironmentWitness,BlumeKohoutZurek2015ObjectiveReality} provide a comprehensive explanation of how quantum states lose their quantum features through interaction with an environment, and how information about the system is redundantly spread across different parts of the environment, giving rise to intersubjective classical records. This program offers a detailed account of how classical states emerge kinematically, but it has thus far not focused on deriving the \emph{effective dynamics} of these states and, in particular, on recovering Newtonian classical mechanics from a given quantum Hamiltonian.

On the other hand, in the domain of continuous measurements it has been shown that, for suitably strong monitoring, one can derive effective classical stochastic equations of motion~\cite{BhattacharyaHabibJacobs2000ClassicalChaos,Carmichael1993QuantumTrajectories,WisemanMilburn2010QuantumMeasurementControl}. However, this approach does not explain classicality in regimes where such continuous monitoring is absent or negligible and typically do not derive classical deterministic evolution of Newtonian mechanics. 

An approach that incorporates elements of the previous two approaches and is most closely related to our present work, is the phase-space quasi-projector framework developed by Omn\`es~\cite{omnes_logical_1989,Omnes1988LogicalReformulationIII,Omnes1997QuantumClassicalProjection} and follow-up studies~\cite{Anastopoulos2002}, as well as the exploration of the classical limit within the consistent history approach~\cite{GellMann1993,Omnes1992ConsistentInterpretations,Omnes1994InterpretationQuantumMechanics,Omnes1990HilbertToCommonSense}. These works associate quasi-projectors in the Hilbert space with large regions of the phase space. One begins with a phase space cell $ C_0 $, chosen sufficiently large and regular (i.e., a “bulky” region with a smooth boundary and no fine-scale structure) so that the localization of a quantum state within it corresponds to a projection onto a subspace of the Hilbert space, represented by a phase-space quasi-projector $ \hat{F}_0 $. The classical Liouville dynamics evolves this cell into $C_t$ over time, with $C_t$ assumed to be sufficiently regular to correspond to a subspace associated to a quasi-projector $\hat{F}_t$. The dynamics of the quantum system under unitary evolution $\hat{U}(t)= \exp({-\frac{i}{\hbar} \hat{H} t})$ is said to exhibit a classical limit if $
\hat{F}(t) \equiv \hat{U}(t) \hat{F}_0 \hat{U}^\dagger(t) = \hat{F}_t 
$. The difference between quantum and classical dynamics is quantified by the error bound:
$  \operatorname{Tr}\bigl|\delta \hat{F}(t)\bigr|
    \leq K\,\epsilon\,\operatorname{Tr}\hat{F}_0,
$ where
$    \delta \hat{F}(t)
    \equiv \hat{F}(t) - \hat{F}_t,
$
$\epsilon$ is the degree of irregularity of the dynamics, and $K$ is a Hamiltonian dependent constant
of order one (with $K = 0$ for at most quadratic Hamiltonians in phase space variables). We therefore conclude that, in order to achieve the classical limit within the approach, (a) the cells $C_t$ must be  sufficiently regular and extended so that one can
associate quantum projectors with the projections onto these cells; (b) the Hamiltonians are well approximated by the quadratic functions of the phase space variables; (c) to apply the method, the boundaries of the cells associated with the projective measurements must be chosen so as to follow the classical dynamics. These leave open the question of how classicality emerges for systems with highly irregular (e.g., chaotic) dynamics and probabilities concentrated in highly localized phase-space regions, or under static,  general POVM measurements. Despite these limitations, the works represent a groundbreaking effort to bridge classicality and quantum mechanics through an innovative methodology. 

Another relevant and complementary research direction is the derivation of classical kinematics and dynamics via epistemic restrictions in quantum mechanics. In particular, starting from classical Liouville mechanics on phase space and supplementing it with uncertainty-like constraints, one can recover the Gaussian sector of quantum theory consisting of Gaussian states, measurements and dynamics~\cite{Bartlett2012}. Related frameworks derive quantum mechanics from classical statistical dynamics supplemented by additional epistemic restrictions (and, in some cases, further ontic structure)~\cite{Budiyono2017}. Our approach proceeds in the opposite direction: starting from unrestricted quantum dynamics, we derive an effective classical phase-space description and Liouville evolution by imposing an operational epistemic restriction in the form of finite-resolution phase-space POVMs.

\section{Summary of the results}

In this work, we provide a comprehensive and rigorous proof of the classical limit of quantum mechanics through coarse-grained measurements, deriving both effective classical kinematics and dynamics, as well as addressing the three questions 1.-3. outlined in the text above. Throughout this work, coarse graining is understood operationally: it represents a finite-resolution measurement of phase space, modeled by a fixed coarse-grained continuous phase-space POVM. The POVM outcomes -- taken as the centers of the POVM elements -- label the points of an \emph{effective phase space} and induce a continuous probability density over it. We demonstrate that when the product of the measurement precisions of position and momentum in phase space (defining a single \enquote{bin}) is much larger than the Planck constant, all quantum states can be effectively described by classical probability distributions in the effective phase space. 

Turning to the dynamics under coarse-grained measurements, we derive an effective equation for the probabilities for outcomes of coarse-grained measurements, containing a Liouville term, a classical non-Liouville term arising from coarse-graining, and quantum corrections. We identify the regime where the corrections to the Liouville evolution can be made arbitrarily small, and the effective dynamics reduce to purely classical Hamiltonian evolution generated by the smoothed Hamiltonian, i.e. the Hamiltonian averaged over the coarse-grained cells. Furthermore, if the cells are sufficiently small that the variation of the quantum Hamiltonian across each cell may be neglected (i.e. if the cell width remains smaller than the derivative of the local radius of curvature of the Hamiltonian in phase space), then the effective dynamics reduce to the classical Hamiltonian flow generated by the classical Hamiltonian $H(x,p)$  obtained from the quantum one $H(\hat{x},\hat{p})$  by replacing phase-space operators with phase-space variables. This condition is indeed satisfied on the macroscopic scale. 

It is worth noting that the quantization procedure, starting from a classical Hamiltonian and passing to a quantum one, is not unique. Because of operator-ordering ambiguities—and, more generally, inequivalent quantization prescriptions—one may arrive at quantum Hamiltonians that differ by commutator terms, that is, by $\hbar$-dependent corrections involving lower-order operators and their derivatives. For our purposes, however, this ambiguity does not affect the classical limit considered here. In the regime of coarse-grained measurements studied in this work, the effective dynamics becomes insensitive to such $\hbar$-suppressed differences, and the resulting description reduces to the same underlying classical Hamiltonian $H(x,p)$. Hence, the methodological loop in Fig.~\ref{fig:1} remains consistent, provided the classical Hamiltonian is understood as the effective Hamiltonian governing the coarse-grained quantum evolution, and provided its variation within each coarse-graining cell is negligible.

Our results concerning the dynamics refer to the evolution of the probability of obtaining a phase-space POVM outcome at a \emph{single time}. To reconstruct the time evolution of this probability, one would therefore need to perform the coarse-grained measurement at various individual times $t$, each time on an independently prepared subensemble and without any prior measurement. The POVMs may be understood either as actually performed, or only counterfactually. In the latter case, they specify which statistics \emph{would} be accessible if the system were probed with a given finite resolution at selected times. This is similar in spirit to coarse-grained histories, but we neither assume nor impose a separate consistency condition. Rather, classicality arises because sufficiently strong coarse-graining renders the relevant observables approximately jointly measurable. In either case, one can estimate the Ehrenfest time, defined as the minimal timescale after which deviations from classical Liouville dynamics become significant.

Unlike the case of measurements performed on independently prepared sub-ensembles at single times, we also consider \emph{repeated} measurements on the same system, thereby studying \emph{multi-time} statistics and the resulting emergence of classical trajectories. The classicality of the full dynamics over an extended period can only be maintained if consecutive coarse-grained measurements are performed before the Ehrenfest time is reached between successive measurements.

The coarse-grained POVMs entering repeated measurements may be read in two ways. In the primary operational reading, they represent actual finite-resolution measurements performed on the systems at chosen times. Alternatively, they may be read counterfactually, as specifying which finite-resolution statistics would be obtained if the systems were probed at those times. This second reading is close in spirit to consistent-histories approach~\cite{Griffiths1984, GellMann1993, Omnes1992ConsistentInterpretations}; however, we do not impose an additional consistency condition, such as decoherence of the decoherence functional. In our approach, the relevant compatibility instead follows from the approximate joint measurability induced by sufficiently coarse graining.

This analysis is applied to two typical scenarios in which classical behavior emerges: (a) macroscopic systems under standard conditions; and (b) microscopic systems under repeated measurements, such as particles in a cloud chamber.

In relation to the closest previous approach~\cite{Omnes1988LogicalReformulationIII,omnes_logical_1989,Omnes1990HilbertToCommonSense,Omnes1992ConsistentInterpretations,Omnes1994InterpretationQuantumMechanics,Omnes1997QuantumClassicalProjection}, we clarify the differences to the previous work as follows:
(i) Omnès’ analysis of phase-space quasi-projectors shows that for quadratic Hamiltonians, where higher-order Moyal terms vanish, quantum and classical evolution essentially coincide; for higher-order potentials, it yields only an upper bound with an unspecified constant of order one. In contrast, we start from a general quantum Hamiltonian and derive explicit conditions under which an effective classical Hamiltonian emerges, while rigorously controlling all errors.
(ii) Omnès uses quasi-projectors whose supports are continually adapted to the evolving classical cells, so that the projectors follow the classical trajectories. We instead work with a single, fixed coarse-grained POVM in phase space, whose elements do not depend on time, the Hamiltonian, or the trajectory—closer to the operational setting of a fixed, finite-resolution measurement device. Furthermore, the time-dependence of the regions in the Omnès approach prevents the treatment of those evolutions that produce irregular regions (e.g., chaotic systems), which is possible in our approach.
(iii) Whereas Omnès restricts attention to deterministic Liouville dynamics in closed systems, our approach provides an explicit dynamical equation for the intermediate regime. This equation contains quantum corrections as well as a diffusion-like contribution generated by coarse graining. In the appropriate limit of large cell sizes, it reduces to a deterministic Liouville equation.

We outline the structure of the remaining parts of the paper. Section~\ref{sec:Definition} defines the classical limit in terms of both kinematics and dynamics, establishing the criteria for transitioning from quantum to classical descriptions. In Section~\ref{sec:Coarsegraining}, we introduce coarse-grained  measurements (continuous POVM) in phase space and demonstrate their compatibility with classical kinematics. Section~\ref{sec:Liouvilledyn} presents the derivation of Liouville dynamics under coarse-grained observations, highlighting the conditions under which classical dynamics emerge, together with the definition of the Ehrenfest time, and the explanation of its role in determining the time scales for the validity of classical approximations. Section~\ref{sec:Trajectories} discusses the emergence of classical trajectories and their dependence on the precision of the measurement and the rate of consecutive measurements. Section~\ref{sec:Parameters} applies these findings to various physical systems, comparing the behavior of macroscopic and microscopic systems under coarse-grained measurements. Finally, Section~\ref{sec:Conclusion} concludes the paper by summarizing the key findings and proposing directions for future research. The appendices provide supporting proofs of the results presented in the main text.

\section{Definition of the classical limit}\label{sec:Definition}

In this section, we formalize the concept of the classical limit. The classical limit has two aspects: a kinematical one, concerning the description of the state at each instant of time, and a dynamical one, concerning its evolution. Kinematically, the classicality is characterized by the condition that, in the limit, the probability distributions associated to the measurements at a fixed time can be understood as marginalizations of an underlying classical phase-space probability distribution. Dynamically, the classicality requires that these phase-space probabilities evolve in accordance with classical laws, specifically through Liouville's equation. For simplicity, we restrict our discussion to a one-dimensional system. However, the conclusions drawn here can readily be extended to systems with higher dimensions.

The kinematical part of the classical limit of quantum mechanics is achieved when, at every given time, the probability distributions associated with the coarse-grained measurements satisfy the following conditions.

\textbf{Classical kinematics}:
\begin{enumerate}
    \item At any time $t$ there exists a non-negative, normalized phase-space probability distribution $P(x,p,t)$, satisfying:
    \begin{equation}
        P(x,p,t) \geq 0, \quad \iint_{-\infty}^\infty dx \, dp \, P(x,p,t) = 1. \label{positive}
    \end{equation}

    \item The probability $P(a \in \mathcal{R},t)$ of obtaining an outcome $a$ in the range $ \mathcal{R}$ at time $t$ is given by:
    \begin{equation}
        P(a \in \mathcal{R},t) = \iint_{\Gamma_{\mathcal{R}}} P(x,p,t) \, dx \, dp, \label{subset}
    \end{equation}
    where $\Gamma_{\mathcal{R}} = \{(x,p) \mid A(x,p,t) \in \mathcal{R}\}$ is the subset of phase space in which the measured physical variable $ A(x,p,t) $ takes values from the range $ \mathcal{R}$.
\end{enumerate}
The two conditions should be understood as criteria under which the probabilities for a given set of measurements admit a fully classical description. In particular, they do not rely on the quantum formalism. More precisely, the left-hand side of Eq.~\eqref{subset} is determined purely operationally by the observed probability of obtaining an outcome within a given range in a measurement, whereas the right-hand side requires the existence of a classical probability distribution whose marginal over the phase-space points corresponding to that outcome reproduces the measured probability.
The conditions directly imply the Kolmogorov axioms of classical probability theory. Specifically, Condition~(\ref{positive}) guarantees non-negativity and normalization of the probability distribution, while Condition~(\ref{subset}) ensures additivity over disjoint sets of outcomes: For any countable sequence of such sets of outcomes $ \mathcal{R}_1, \mathcal{R}_2, \dots $ (i.e., $ \mathcal{R}_i \cap \mathcal{R}_j = \emptyset $ for $ i \neq j $), the probability of finding an outcome in their union equals the sum of the probabilities for finding it in individual sets, i.e. 
$
P\left( \bigcup_{i=1}^\infty \mathcal{R}_i \right) = \iint_{\cup_{i=1}^\infty\Gamma_{\mathcal{R}_i}} dx dp \; P(x,p,t) = \sum_i \iint_{\Gamma_{\mathcal{R}_i}} dx dp \; P(x,p,t)=\sum_{i=1}^\infty P(\mathcal{R}_i),$
where $ \Gamma_{\mathcal{R}_1},\Gamma_{\mathcal{R}_2}, \dots $ ($ \Gamma_{\mathcal{R}_i} \cap \Gamma_{\mathcal{R}_j} = \emptyset $ for $ i \neq j $) are disjoint regions in the phase space, corresponding to the mutually exclusive sets of outcomes. 

Although these conditions highlight a key distinction between the kinematics of quantum and classical mechanics, it is known that quasi-probability distributions are often used in quantum theory to calculate measurement probabilities. However, none of these distributions satisfy both classical conditions simultaneously. For example, the Wigner function satisfies Condition~(\ref{subset}) but violates Condition~(\ref{positive}) since it can take negative values, while the Husimi $Q$-function satisfies Condition~(\ref{positive}) but fails to fulfill Condition~(\ref{subset}) for general quantum measurements. This failure reflects the well-known incompatibility of quantum mechanics with a classical, non-contextual hidden variable model. Nevertheless, as we will show, our framework allows these classical conditions to be satisfied once the measurements on the system are restricted to sufficiently coarse-grained observables. We will later show that, for this restricted class of measurements, the $Q$-function is essential for furnishing a classical probability distribution in the sense of Eqs.~\eqref{positive} and~\eqref{subset}. We now proceed to define the dynamical classical limit.

\textbf{Classical (deterministic) dynamics} 
for a closed system is defined to hold when the probability distribution $P(x,p,t)$, satisfying Eqs.~(1) and~(2), evolves according to Liouville’s equation in phase space:
\begin{equation}
    \frac{\partial P}{\partial t} = -\{P, H\}, \label{lioville}
\end{equation}
where $\{ \cdot \, , \cdot\}$ denotes the Poisson bracket, and $H(x,p,t)$ is the classical Hamiltonian governing the system.

 We will next define coarse-grained measurements in phase space. They will be modeled by a continuous phase-space POVM, whose continuous set of outcomes defines an \textit{``effective phase space''}. We then show that, for sufficiently large coarse graining, the probability distribution on this effective phase space satisfies the kinematic and dynamic conditions of classicality. 

\section{The coarse-grained phase-space measurements}
\label{sec:Coarsegraining}

In this section we introduce a family of coarse-grained phase-space measurements obtained by Gaussian smearing of the canonical coherent-state POVM, analogous to what was discussed for spin systems in Ref.~\cite{kofler2007}. This construction yields a positive, normalized phase-space probability density that reduces to the Husimi $Q$-function in the zero-smearing limit. We further show that, in the limit of large smearing, coarse-grained observables constructed from this POVM become approximately commuting.

\begin{definition}[\textbf{Coarse-grained phase-space POVM}]\label{def}
We consider the coarse-grained phase-space POVM defined by
\begin{align}
  \hat{P}_{\boldsymbol\Delta}(x,p)
  := \frac{1}{2\pi\hbar}\!&\iint_{\mathbb{R}^2}\! dx'\,dp'\;
     G_{\boldsymbol\Delta}(x-x',p-p')\,\ket{x',p'}\!\bra{x',p'},\label{eq:POVM_definition}
\end{align}
where $\ket{x \,p}=e^{\frac{i}{\hbar}(x \hat{p} - p \hat{x})}\ket{0}$ are coherent states centered in the phase-space point $(x,p)$ with uncertainties $\sigma_x$, $\sigma_p$ satisfying $\sigma_x \sigma_p =\frac{\hbar}{2}$.  $G_{\boldsymbol\Delta}(x,p)$ is a Gaussian distribution centered in the origin, with standard deviations $\Delta_x \sigma_x$ and $\Delta_p \sigma_p$, i.e. $G_{\boldsymbol\Delta}(x,p)
  := \frac{1}{2\pi\,\Delta_x\Delta_p\,\sigma_x\sigma_p}\,
     \exp\!\left(-\frac{x^2}{2\Delta_x^2\sigma_x^2}-\frac{p^2}{2\Delta_p^2\sigma_p^2}\right)$. We will refer to $\boldsymbol{\Delta}=(\Delta_x,\Delta_p)$ as the dimensionless coarse-graining parameters.
\end{definition}

It is straightforward to check that Eq.~\eqref{eq:POVM_definition} defines a POVM on $\mathbb{R}^2$ with measure $dx \,dp$, i.e. $\hat P_{\boldsymbol \Delta}\geq 0$ and $\int_{\mathbb{R}^2} dx \,dp\,\hat P_{\boldsymbol \Delta}(x,p)=\mathbb{1}$.

\begin{figure}[htb!]
\centering
    % Panel a
    \begin{subfigure}[b]{0.45\textwidth}
        \centering
        \includegraphics[width=\textwidth]{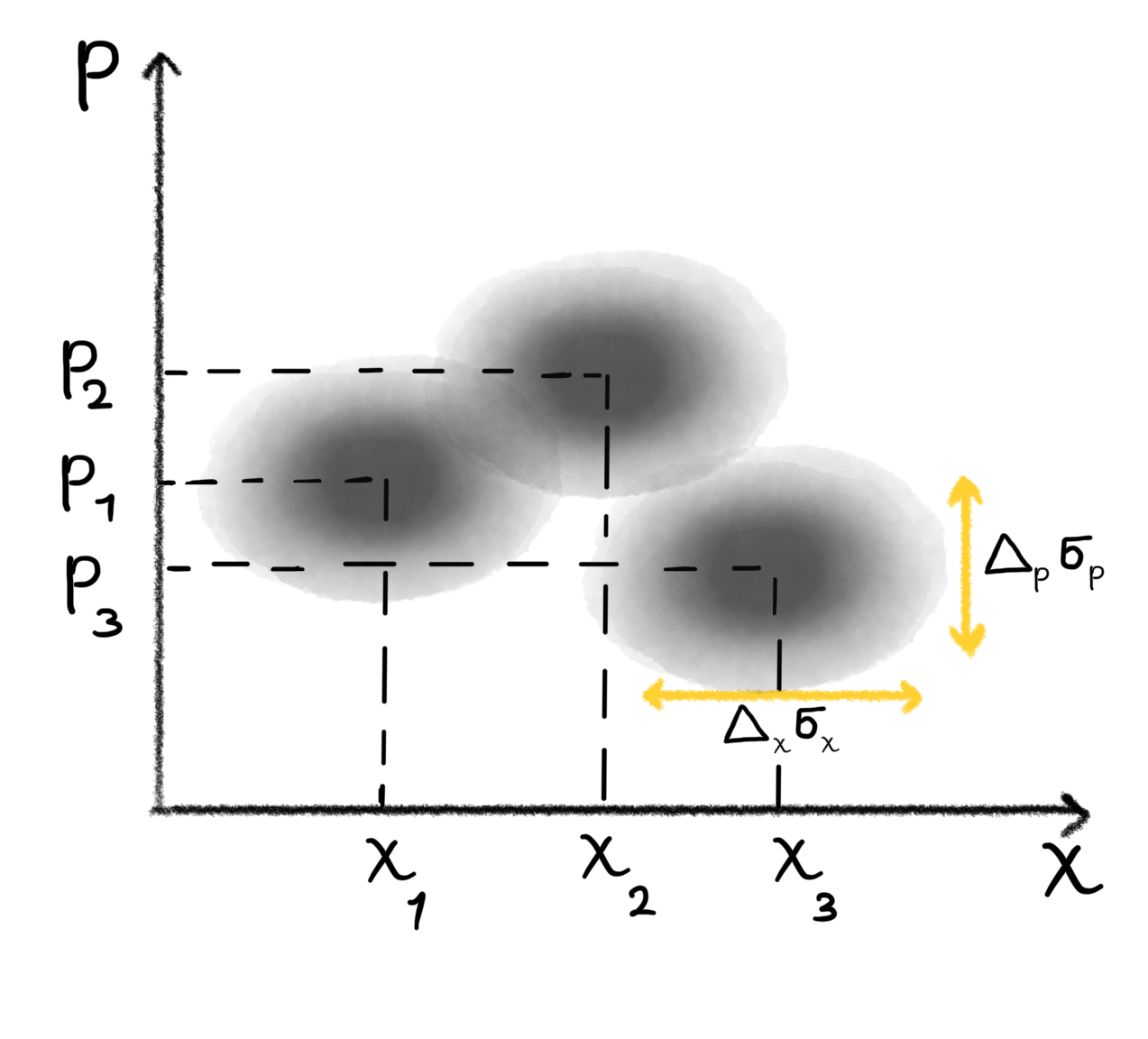} % Replace with your image
    \end{subfigure}
    \hfill
    % Panel b
    \begin{subfigure}[b]{0.45\textwidth}
        \centering
        \includegraphics[width=\textwidth]{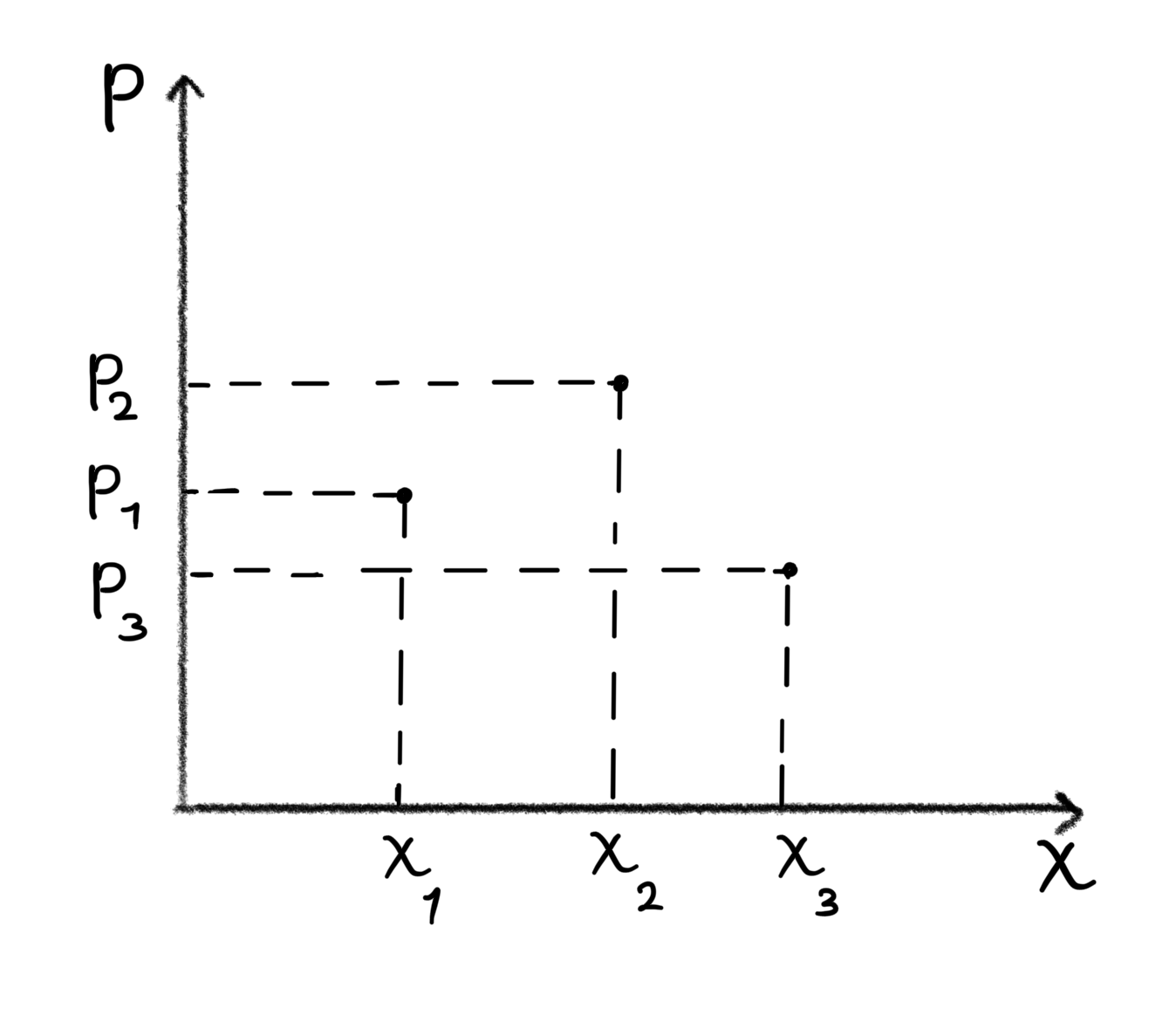} % Replace with your image
       \end{subfigure}
\caption{Comparison between the original and the effective phase space. Left: POVM elements are associated with a Gaussian-weighted integral over the original phase space, centered at the outcome $(x,p)$. Right: to each outcome $(x,p)$ one associates a single point $(x,p)$ in the effective phase space. In this sense, the effective phase space is a smoothed version of the original phase space.}
\label{rescaling}
\end{figure}

Using the coarse-grained phase-space POVM, we construct an \textit{effective phase space} in which each point $(x,p)$ is represented by the center of the corresponding POVM element, or equivalently by the outcome of the coarse-grained phase-space measurement as illustrated in Fig.~\ref{rescaling}. Hence, the phase points of the effective phase space are determined by Gaussian-weighted neighborhoods of points in the original phase space. In what follows, we define the probability distribution on the effective phase space and show that any observables built from the POVM elements, become approximately commuting once the effective coarse-graining cell area is large compared Planck’s constant.

\paragraph{Associated probability density.}
For a state $\hat{\rho}(t)$, the probability density to observe the
outcome $(x,p)$ of the coarse-grained POVM at time $t$ is given by
\begin{equation}
  p_{\boldsymbol\Delta}(x,p,t)
  := \mathrm{Tr}\!\big[\hat{P}_{\boldsymbol\Delta}(x,p)\,\hat\rho(t)\big]
  = \iint dx'\,dp'\;G_{\boldsymbol\Delta}(x-x',p-p')\,Q(x',p',t),
\label{eq:probabilityDistribution}
\end{equation}
where 
\begin{equation}
  Q(x,p,t) := \frac{1}{2\pi\hbar}\,\braket{x,p|\hat\rho(t)|x,p}
\end{equation}
is the Husimi function of the state. 

Since $\hat P_{\boldsymbol \Delta}$ is a POVM, $p_{\boldsymbol\Delta}$
satisfies the non-negativity and normalization conditions of Eq.~\eqref{positive}. In
the limit $\Delta_x,\Delta_p\to 0$, $G_{\boldsymbol\Delta}$ approaches a
delta-function and $p_{\boldsymbol\Delta}(x,p,t)$ approaches the Husimi function $Q(x,p,t)$. 

Accordingly, one can define coarse-grained observables as

\begin{definition}[Coarse-grained observables]
Given a real function $A:\mathbb{R}^2\!\to\!\mathbb{R}$, we define the associated \emph{coarse-grained observable}
\begin{equation}
  \hat A_{\boldsymbol\Delta}
  := \iint_{\mathbb{R}^2}\! A(x,p)\,\hat{P}_{\boldsymbol\Delta}(x,p)\,dx\,dp
  \;=\;\frac{1}{2\pi\hbar}\iint_{\mathbb{R}^2}\!
  A_{\boldsymbol\Delta}(x,p)\,\ket{x,p}\!\bra{x,p}\,dx\,dp,
  \label{eq:CGo}
\end{equation}
where $A_{\boldsymbol\Delta}=G_{\boldsymbol\Delta}*A$ denotes Gaussian
coarse-graining of the function $A$ by the kernel $G_{\boldsymbol\Delta}$ from
Def.~\ref{def}.
\end{definition}

We are now ready to derive our first result: with sufficient coarse graining, observables in the effective phase space become approximately commuting.

\begin{statement}
\label{prop:Decay}
The coarse-grained observables $\hat A_{\boldsymbol\Delta}$ and
$\hat B_{\boldsymbol\Delta}$ become approximately commuting as the
coarse-graining widths $\Delta_x,\Delta_p$ are taken to be much larger than unity. More precisely,

\begin{equation}
  \big\|\,[\hat A_{\boldsymbol\Delta},\hat B_{\boldsymbol\Delta}]\,\big\|
  \;\le\;
  \,
  \frac{\|A\|_{L^1_\hbar}\,\|B\|_{L^1_\hbar}}
       {\big(1+\tfrac{\Delta_x^2}{4}\big)^{3/2}
        \big(1+\tfrac{\Delta_p^2}{4}\big)^{3/2}}.
  \label{eq:MainBound-L1-dimensionless}
\end{equation}

where we introduce
the dimensionless symplectic $L^1_\hbar$ norm
\begin{equation}
  \|A\|_{L^1_\hbar}
  :=\frac{1}{2\pi\hbar}\iint_{\mathbb{R}^2} |A(x,p)|\,dx\,dp.
\end{equation}
This estimate shows that the
operator norm of the commutator $[\hat A_{\boldsymbol\Delta},
\hat B_{\boldsymbol\Delta}]$ vanishes in the limit $\max\{\Delta_x,\Delta_p\}\!\to\!\infty$, with asymptotic decay
\begin{equation}
    \big\|\,[\hat A_{\boldsymbol\Delta},\hat B_{\boldsymbol\Delta}] \big\|=\mathcal{O}\big(\Delta^{-3}_x,\Delta^{-3}_p\big).
\end{equation}
In other words, sufficiently strong, in general anisotropic, coarse-graining in
phase space suppresses non-commutativity and renders all such
coarse-grained observables jointly approximately measurable.
\end{statement}

\begin{proof}
The proof consists in rewriting the coarse-grained observables $\hat A_{\boldsymbol \Delta}$ and $B_{\boldsymbol \Delta}$ in Weyl form, isolating the contribution of the coarse-graining through the convolution theorem. Using the Weyl relations leads to a bound on $\big\|\,[\hat A_{\boldsymbol\Delta},\hat B_{\boldsymbol\Delta}]\,\big\|$ obtained by solving an explicit integration. Details are in Appendix~\ref{app:statement1}.
\end{proof}

Note that this bound is well defined only when the $||\cdot||_{L_\hbar^1}$ are finite. Here we assume the physically motivated scenarios where the observation region in phase-space is limited to a bounded region $K \subset \mathbb{R}^2$ that contains (up to negligible tails) the support of the state. In this scenario we can always replace all observables with their cut-off versions in \(K\). Equivalently, in Eq.~\eqref{eq:CGo} the phase-space integral is restricted to the bounded observation region \(K\subset\mathbb{R}^2\), rather than taken over the whole of \(\mathbb{R}^2\). The resulting cut-off observables have finite \(L^1_\hbar\)-norms.

Statement 1 states that, under sufficiently coarse-grained resolution of the measurement apparatus, the kinematics of any quantum state becomes effectively classical, since all observables approximately commute. Note that the statement makes no reference to the initial coherent state with $\sigma_x$ and $\sigma_p$ from which the POVM is constructed. This means that, for all states with support within the bounded phase-space region $K$, Statement 1 specifies an entire \emph{class of POVMs} of type~\eqref{eq:POVM_definition} that all lead to classical kinematics, provided that $\Delta_x$ and $\Delta_p$ are sufficiently large.

\section{The limit of Liouville's dynamics}

\label{sec:Liouvilledyn}

We now quantify when the coarse-grained phase-space statistics $p_{\boldsymbol\Delta}(x,p,t)$ evolves essentially classically, \emph{i.e.}\ under Liouville transport, and how coarse graining suppresses the residual quantum corrections.

Recall that $p_{\boldsymbol\Delta}$ was defined in Eq.~\eqref{eq:probabilityDistribution} by convolving the Husimi function $Q$ with a Gaussian. To study how $p_{\boldsymbol\Delta}$ evolves in time, it is convenient to work instead with the Wigner function $W$, whose dynamics are given explicitly. We rewrite $p_{\boldsymbol\Delta}$ in terms of $W$, use the known evolution equation for $W$, and then recover $p_{\boldsymbol\Delta}$ by applying the same Gaussian convolution.
Since $Q$ is itself a Gaussian smoothing of $W$, the combined smoothing from $W$ to $p_{\boldsymbol\Delta}$ is again Gaussian,
so that 
\begin{equation}\label{eq:pDelta_smoothing}
p_{\boldsymbol\Delta}=G_{
\left(\sqrt{\Delta_x^2+1},\sqrt{\Delta_p^2+1}\right)} * W=S_{\boldsymbol\Delta}W, \qquad 
 S_{\boldsymbol\Delta}:=\exp\!\Big(\tfrac{\ell_x^2}{2}\partial_x^2+\tfrac{\ell_p^2}{2}\partial_p^2\Big),
\end{equation}
where
\begin{equation}\label{eq:coarse-grained-variances}
\ell_x^2:=\sigma_x^2(\Delta_x^2+1), \qquad \ell_p^2:=\sigma_p^2(\Delta_p^2+1)
\end{equation}
are coarse-grained variances. The large coarse-graining limit corresponds to $\ell_x,\ell_p\to\infty$.
\begin{statement}\label{st:cg_exact_min}
Let $\hat H$ have Weyl symbol $H(x,p)=\int dy\, e^{-i p y/\hbar}\,\Big\langle x+\tfrac{y}{2}\Big|\hat H\Big|x-\tfrac{y}{2}\Big\rangle$ and define the smoothed Hamiltonian
$H_{\boldsymbol\Delta}:=S_{\boldsymbol\Delta}H$.
Then $p_{\boldsymbol\Delta}$ satisfies the exact closed equation
\begin{equation}\label{eq:cg_exact_clean}
\partial_t p_{\boldsymbol\Delta}
=
L_{H_{\boldsymbol\Delta}}p_{\boldsymbol\Delta}
+\mathcal D_{\boldsymbol\Delta}p_{\boldsymbol\Delta}
+\mathcal{R}_{\boldsymbol\Delta}p_{\boldsymbol\Delta}.
\end{equation}
The three terms are defined by 
\begin{align}
L_{H_{\boldsymbol\Delta}}f
&:=\{H_{\boldsymbol\Delta},f\},
\label{eq:L_def_min}\\[2mm]
\mathcal D_{\boldsymbol\Delta}f
&:=S_{\boldsymbol\Delta}\!\left(\{H,S_{\boldsymbol\Delta}^{-1}f\}\right)
-\{H_{\boldsymbol\Delta},f\},
\label{eq:DDelta_def_min}\\[2mm]
\mathcal{R}_{\boldsymbol\Delta} f
&:=S_{\boldsymbol\Delta}\!\left(\frac{2}{\hbar}\,H\,\sin\!\Big(\frac{\hbar}{2}\Lambda\Big)\,S_{\boldsymbol\Delta}^{-1}f
-\{H,S_{\boldsymbol\Delta}^{-1}f\}\right),
\label{eq:RDelta_def_min}
\end{align}
where $\Lambda:=\overleftarrow{\partial}_x\,\overrightarrow{\partial}_p-\overleftarrow{\partial}_p\,\overrightarrow{\partial}_x$.
\end{statement}
\begin{proof}
The proof is straightforward. The Wigner function $W$ satisfies the Moyal equation. Applying $S_{\boldsymbol\Delta}$ and using the identity $W=S_{\boldsymbol\Delta}^{-1}p_{\boldsymbol\Delta}$ gives an exact evolution equation for $p_{\boldsymbol\Delta}$ in which the Moyal generator is conjugated by the smoothing map. We group the $\hbar$-dependent terms in $\mathcal{R}_{\boldsymbol\Delta}$ and the second and higher order derivatives in $\mathcal{D}_{\boldsymbol\Delta}$.

\end{proof}

The operator $\mathcal{R}_{\boldsymbol\Delta}$ is the genuinely quantum contribution: it is the Moyal remainder beyond the Poisson bracket, transported through the smoothing map. By contrast, $\mathcal D_{\boldsymbol\Delta}$ is a classical correction to Liouville evolution, arising because coarse graining and Liouville transport do not commute.

\begin{statement}\label{st:duhamel}Fix a bounded observation region $K \subset \mathbb{R}^2$.
Let $p_{\boldsymbol \Delta}(t)$ be the coarse-grained density, and let
$p_{\boldsymbol \Delta}^{\mathrm{cl}}(t)$ denote the classical reference evolving by Liouville transport
generated by the coarse grained Hamiltonian $H_{\boldsymbol \Delta}$,
\begin{equation} \label{eq:pclassical}
\partial_t p_{\boldsymbol \Delta}^{\mathrm{cl}} = L_{H_{\boldsymbol \Delta}}\,p_{\boldsymbol \Delta}^{\mathrm{cl}},
\qquad
p_{\boldsymbol \Delta}^{\mathrm{cl}}(0)=p_{\boldsymbol \Delta}(0).
\end{equation}
Then the deviation of $p_{\boldsymbol\Delta}(t)$ from Liouville evolution on $K$ is bounded by

\begin{align} \label{eq:time_evolution_bound}
\|p_{\boldsymbol \Delta}(t) - p_{\boldsymbol \Delta}^\text{cl}(t)\|_{L^\infty(K)}
\leq\; \int_0^t \|\mathcal D_{\boldsymbol\Delta}p_{\boldsymbol\Delta}(s)\|_{L^\infty(K_{t-s})}\,ds
+ \int_0^t \|\mathcal R_{\boldsymbol\Delta}p_{\boldsymbol\Delta}(s)\|_{L^\infty(K_{t-s})}\,ds,
\end{align}
with $K_t:=\varphi^{-t}(K)$ being the region spanned by the backward evolution, defined by the
Hamiltonian flow $e^{t L_{H_{\boldsymbol \Delta}}}f = f \circ \varphi^{-t}$, where $f$ is a function on a phase space.
\end{statement}

\begin{proof}
Consider the difference $p_{\boldsymbol \Delta}-p_{\boldsymbol \Delta}^{\mathrm{cl}}$ and use the fact that
$p_{\boldsymbol \Delta}$ satisfies the exact decomposition
$\partial_t p_{\boldsymbol \Delta}=L_{H_{\boldsymbol \Delta}} p_{\boldsymbol \Delta}+\mathcal D_{\boldsymbol\Delta}p_{\boldsymbol\Delta}+\mathcal R_{\boldsymbol\Delta}p_{\boldsymbol\Delta}$,
while $p_{\boldsymbol \Delta}^{\mathrm{cl}}$ evolves by $\partial_t p_{\boldsymbol \Delta}^{\mathrm{cl}}=L_{H_{\boldsymbol \Delta}} p_{\boldsymbol \Delta}^{\mathrm{cl}}$.
Subtracting the two equations gives a linear transport equation for the difference: $\partial_t (p_{\boldsymbol \Delta}-p_{\boldsymbol \Delta}^\text{cl}) - L_{H_{\boldsymbol \Delta}}(p_{\boldsymbol \Delta}-p_{\boldsymbol \Delta}^\text{cl}) = \mathcal{D}_{\boldsymbol\Delta}p_{\boldsymbol\Delta}+\mathcal R_{\boldsymbol\Delta}p_{\boldsymbol\Delta}$. Solving this inhomogeneous transport equation by Duhamel's formula expresses the difference at time $t$ as a time integral
of the source term transported backward along the Hamiltonian flow of $H_{\boldsymbol \Delta}$. Taking the $L^\infty$ norm on $K$ and using that
Liouville transport acts by composition with the flow, hence preserves $L^\infty$ norms up to the transported domain $K_{t-s}$,
yields the stated bound after bringing the norm inside the integral and separating the two contributions. Details are
given in Appendix~\ref{app:statement3}.
\end{proof}

In the following statements we quantify the upper bounds for the operators appearing on the RHS of Eq.~\eqref{eq:time_evolution_bound}.
\begin{statement}\label{st:R_bound_poly}
Let $H(x,p)=\frac{p^{2}}{2m}+V(x)$ and fix a bounded observation region $K\subset\mathbb R^{2}$. Then
\begin{equation}\label{eq:R_bound}
    \|\mathcal{R}_{\boldsymbol\Delta}p_{\boldsymbol\Delta}\|_{L^\infty(K)}
\le
\frac{1}{2\pi\,\ell_x\ell_p}
\sum_{n=1}^{\infty}\frac{\kappa_{2n+1}}{(2n+1)!}\Big(\frac{\hbar}{2}\Big)^{2n}\,
\frac{1}{\ell_p^{2n+1}}\;
\big\|(\partial_x^{2n+1}V)\,W\big\|_{L^1(\mathbb R^{2})},
\end{equation}
where $\kappa_k:=\sup_{u\in\mathbb R}\big|\mathrm{He}_k(u)\big|e^{-u^{2}/2}$, and
$\mathrm{He}_k$ are the probabilists' Hermite polynomials. Therefore
\begin{equation}
\|\mathcal{R}_{\boldsymbol\Delta}p_{\boldsymbol\Delta}\|_{L^\infty(K)}\to 0
\qquad\text{as}\qquad
\quad\ell_x\ell_p\rightarrow\infty,
\end{equation}
as long as the series in Eq.~\eqref{eq:R_bound} is convergent and $\big\|(\partial_x^{2n+1}V)\,W\big\|_{L^1(\mathbb R^{2})}$ is finite, which holds true under the standard assumptions of well-behaved dynamics and $W$ being sufficiently smooth and rapidly decaying. 
\end{statement}

\begin{proof}
Expand the sine in Eq.~\eqref{eq:RDelta_def_min} and subtract the Poisson-bracket term, so that
$\mathcal{R}_{\boldsymbol\Delta}p_{\boldsymbol\Delta}$ is given by an infinite odd-derivative remainder.
Using that $S_{\boldsymbol\Delta}$ is Gaussian smoothing, one rewrites each term by moving the $p$-derivatives onto the
Gaussian kernel (justified by the assumed decay of $W$). Young's inequality then bounds the $L^\infty(K)$-norm by the
product of an $L^\infty$-norm of a $p$-derivative of the Gaussian and the $L^1$-norm of
$(\partial_x^{2n+1}V)W$. Finally, Gaussian derivatives are Hermite polynomials times the Gaussian, which yields the stated
series estimate with the explicit factor $\ell_p^{-(2n+1)}$ and the prefactor
$(2\pi\,\ell_x\ell_p)^{-1}$. The limit $\ell_x\ell_p\to\infty$ follows immediately. Details are
given in Appendix~\ref{app:statement4}.
\end{proof}

\begin{statement}\label{st:D_bound}
Let $H(x,p)=\frac{p^{2}}{2m}+V(x)$, and let $K\subset\mathbb R^{2}$ be bounded. Define
\begin{equation}
V_{\boldsymbol \Delta}:=S_{\boldsymbol \Delta} V.
\end{equation}
Then
\begin{equation}\label{eq:D_bound_maintext}
\|\mathcal D_{\boldsymbol\Delta}p_{\boldsymbol\Delta}\|_{L^\infty(K)}
\le
\frac{1}{2\pi}\|W\|_{L^1(\mathbb R^2)}
\left(
\frac{\kappa_1}{\ell_x\ell_p^2}
\sum_{n\ge1}\frac{\kappa_n}{n!}\,\ell_x^{\,n}\,
\|V_{\boldsymbol \Delta}^{(n+1)}\|_{L^\infty(K)}
+
\frac{e^{-1}}{m\ell_x^2}
\right),
\end{equation}
where
\begin{equation}
\kappa_n:=\sup_{u\in\mathbb R} |\mathrm{He}_n(u)|e^{-u^2/2}.
\end{equation}

Assume moreover that the smoothed curvature varies on a characteristic length $L_V(K)$, in the sense that
\begin{equation}
\|V_{\boldsymbol \Delta}^{(n+1)}\|_{L^\infty(K)}
\lesssim
\frac{\|V_{\boldsymbol \Delta}^{(2)}\|_{L^\infty(K)}}{L_V(K)^{\,n-1}},
\qquad n > 1.
\end{equation}
Then
\begin{equation}\label{eq:D_bound_simplified}
\|\mathcal D_{\boldsymbol\Delta}p_{\boldsymbol\Delta}\|_{L^\infty(K)}
\lesssim
\frac{1}{2\pi}\|W\|_{L^1(\mathbb R^2)}
\left(
\frac{\kappa_1}{\ell_p^2}\,
\|V_{\boldsymbol \Delta}^{(2)}\|_{L^\infty(K)}
\sum_{n\ge1}\frac{\kappa_n}{n!}
\left(\frac{\ell_x}{L_V(K)}\right)^{n-1}
+
\frac{e^{-1}}{m\ell_x^2}
\right).
\end{equation}
Moreover, the series
\begin{equation}
\sum_{n\ge1}\frac{\kappa_n}{n!}
\left(\frac{\ell_x}{L_V(K)}\right)^{n-1}
\end{equation}
is absolutely convergent for every $\ell_x/L_V(K)$ (see \ref{app:absolute_convergence}). In particular, when
\begin{equation}
\frac{\ell_x}{L_V(K)}\ll1,
\end{equation}
the terms with $n\ge2$ are higher-order corrections, so the potential contribution is governed to leading order by
\begin{equation}
\frac{\|V_{\boldsymbol \Delta}^{(2)}\|_{L^\infty(K)}}{\ell_p^2}.
\end{equation}
\end{statement}
\begin{proof}
The operator $\mathcal D_{\boldsymbol\Delta}$ is the commutator between Liouville transport and Gaussian coarse graining. It therefore splits naturally into a kinetic contribution, coming from the transport term $(p/m)\partial_x$, and a potential contribution, coming from the force term $V'(x)\partial_p$.

The kinetic part is controlled by the fact that Gaussian smoothing in momentum does not exactly commute with multiplication by $p$. This produces a term involving the mixed derivative $\partial_{xp}p_{\boldsymbol\Delta}$. Since $p_{\boldsymbol\Delta}$ is a Gaussian convolution of $W$, Young's inequality and the explicit derivative bounds for the Gaussian kernel give the second term in Eq.~\eqref{eq:D_bound_maintext}, which is suppressed as $1/\ell_x^2$.

The potential part comes from the fact that Gaussian smoothing in position does not exactly commute with multiplication by the force $V'(x)$. Expanding this commutator gives a series involving higher spatial derivatives of the smoothed force and corresponding derivatives of $p_{\boldsymbol\Delta}$. Estimating the latter again by Gaussian-kernel bounds yields the first term in Eq.~\eqref{eq:D_bound_maintext}. Full details are given in Appendix~\ref{app:statement5}.
\end{proof}

\subsection{Ehrenfest time scale}

\label{subsec:Ehrenfest}

In Statement~\ref{st:duhamel} we showed that the difference between the exact evolution of $p_{\boldsymbol\Delta}$ and its classical approximation $p_{\boldsymbol\Delta}^\text{cl}$ is bounded by a correction term. Here we estimate what is the maximum time under which the two evolved distributions are indistinguishable, defining the \emph{Ehrenfest time} for the coarse-grained description. This motivates defining the (coarse-grained) Ehrenfest time as the largest time up to which the Liouville approximation stays within
a tolerance $\delta>0$:
\begin{equation}\label{eq:tE-quant-def}
t_{\mathrm E}
:=
\sup\Big\{T\ge0:\;
\|p_{\boldsymbol\Delta}(t)-p_{\boldsymbol\Delta}^\text{cl}(t)\|_{L^\infty(K)}\le\delta
\ \ \forall\,t\in[0,T]\Big\}.
\end{equation}
For any fixed $T>0$, define
\begin{equation}\label{eq:GammaDelta-T}
\Gamma_{\boldsymbol\Delta}(T)
:=
\sup_{0\le s\le T} \left(
\|\mathcal D_{\boldsymbol\Delta}p_{\boldsymbol\Delta}(s)\|_{L^\infty(K_{T-s})}
+
\|\mathcal R_{\boldsymbol\Delta}p_{\boldsymbol\Delta}(s)\|_{L^\infty(K_{T-s})}
\right).
\end{equation}
Then Eq.~\eqref{eq:time_evolution_bound} implies
\begin{equation}\label{eq:delta-p-linear}
\|p_{\boldsymbol\Delta}(T)-p^\text{cl}_{\boldsymbol\Delta}(T)\|_{L^\infty(K)}
\le
T\,\Gamma_{\boldsymbol\Delta}(T).
\end{equation}
Therefore, a sufficient condition for the Liouville approximation to remain within the tolerance $\delta$ up to time $T$ is
\begin{equation}
T\,\Gamma_{\boldsymbol\Delta}(T)\le \delta.
\end{equation}
Equivalently, whenever this condition holds, one has
\begin{equation}\label{eq:tE-lowerbound}
t_{\mathrm E}\ge T.
\end{equation}

In particular, if $T$ is small and $\Gamma_{\boldsymbol\Delta}(T)=\Gamma_{\boldsymbol\Delta}(0)+O(T)$, then to leading order
\begin{equation}\label{eq:tE-quant-estimate}
t_{\mathrm E}\gtrsim \frac{\delta}{\Gamma_{\boldsymbol\Delta}(0)}.
\end{equation}

The correction rate $\Gamma_{\boldsymbol\Delta}$ receives two distinct contributions. The first, $\mathcal R_{\boldsymbol\Delta}$, is genuinely quantum and comes from the Moyal terms beyond Poisson order. The second, $\mathcal D_{\boldsymbol\Delta}$, is classical and arises because the coarse-graining map does not exactly commute with Liouville evolution. Using Statements~\ref{st:R_bound_poly} and~\ref{st:D_bound}, in the approximation of small Ehrenfest time, we therefore find
\begin{equation}\label{eq:ehrenfest}
        t_E > \frac{\delta}{\Gamma_{\mathcal{R}}(0) + \Gamma_{\mathcal D}(0)}
\end{equation}
with
\begin{align}
    \Gamma_{\mathcal{R}}(t) &= \sup_{0 \leq s \leq t} \frac{1}{2\pi\,\ell_x\ell_p}
\sum_{n=1}^{\infty}\frac{\kappa_{2n+1}}{(2n+1)!}\Big(\frac{\hbar}{2}\Big)^{2n}\,
\frac{1}{\ell_p^{2n+1}}\;
\big\|(\partial_x^{2n+1}V)\,W(s)\big\|_{L^1(\mathbb R^{2})},
\\
    \Gamma_{\mathcal{D}}(t) &= \sup_{0 \leq s \leq t}\frac{1}{2\pi}\|W(s)\|_{L^1(\mathbb R^2)}\Big(\frac{\kappa_1}{\ell_x\,\ell_p^2}\,
\|V_{\boldsymbol \Delta}^{(2)}\|_{L^\infty(K_{t-s})}
\sum_{n\ge1}\frac{\kappa_n}{n!}\left(\frac{\ell_x}{L_V(K_{t-s})}\right)^{n-1}
+
\frac{e^{-1}}{m\,\ell_x^2}\Big). \label{gammaD}
\end{align}

In the simplest case of quadratic Hamiltonians, the Moyal bracket truncates to the Poisson bracket, so the quantum remainder vanishes identically: $\mathcal R_{\boldsymbol\Delta}=0$ and hence $\Gamma_{\mathcal R}=0$. The only remaining correction is therefore $\mathcal D_{\boldsymbol\Delta}$, which reflects the fact that Gaussian coarse graining and Liouville transport need not commute even for exactly quadratic dynamics. This contribution is suppressed in the strong coarse-graining regime $\ell_x,\ell_p\to\infty$.

In the case of free evolution, we obtain
\begin{align}
    \Gamma_{\mathcal R}(t) &= 0 \\
    \Gamma_{\mathcal D}(t) &= \sup_{0\leq s \leq t} \frac{1}{2\pi}\|W(s)\|_{L^1(\mathbb R^2)}\frac{e^{-1}}{m\ell_x^2}.
    \label{free}
\end{align}
In the chaotic regime, the classical flow stretches and folds phase-space volumes, so that the backward-transported observation region $K_t=\varphi^{-t}(K)$
typically grows rapidly along unstable directions. Physically, this reflects the exponential sensitivity to initial conditions encoded by positive Lyapunov exponents. Since the correction terms in Eq.~\eqref{eq:time_evolution_bound} are controlled on $K_t$, one expects their magnitude to inherit a corresponding growth in time. The Duhamel estimate then suggests that, in chaotic systems, the deviation from Liouville evolution may accumulate exponentially fast, leading to the familiar logarithmic scaling of the Ehrenfest time.

To illustrate this heuristically, suppose that $\|V_{\boldsymbol \Delta}^{(2)}\|_{L^\infty(K_t)}$ grows exponentially in time and that, correspondingly,
\[
\Gamma_{\mathcal D}(t)\lesssim C(\ell_x,\ell_p)e^{\lambda t},
\]
where $\lambda$ is a characteristic Lyapunov exponent and $C(\ell_x,\ell_p)$ denotes the prefactor multiplying $\|V_{\boldsymbol \Delta}^{(2)}\|_{L^\infty(K_t)}$ in Eq.~\eqref{gammaD}. Then Eq.~\eqref{eq:time_evolution_bound} yields
\[
\|p_{\boldsymbol\Delta}(t)-p_{\boldsymbol\Delta}^{\mathrm{cl}}(t)\|_{L^\infty(K)}
\lesssim
\int_0^t C(\ell_x,\ell_p)e^{\lambda s}\,ds
=
\frac{C(\ell_x,\ell_p)}{\lambda}(e^{\lambda t}-1).
\]
Under this growth assumption, one therefore obtains the logarithmic lower-bound estimate
\begin{equation}
  t_E \gtrsim \frac{1}{\lambda}\ln\Bigl(1+\frac{\lambda\delta}{C(\ell_x,\ell_p)}\Bigr).  
\end{equation}

We find that, for chaotic systems, the Ehrenfest time is bounded by a logarithmic correction to the Lyapunov time. Nevertheless, the Ehrenfest time can be much larger than the Lyapunov time, due to the scaling in $l_x$ and $l_p$ inside the logarithm.
\subsection{Closing the circle}

We can now turn to the question of when quantization yields a quantum Hamiltonian whose classicalization through coarse-grained measurements reproduces the original classical Hamiltonian. Equivalently, we ask under what conditions the loop in Fig.~\ref{fig:1} closes. Since $H_{\boldsymbol\Delta}$ is obtained by averaging $H$ over a phase-space cell of widths $(\ell_x,\ell_p)$, this requires that such averaging does not significantly modify the classical flow generated by $H$. In other words, the coarse-graining must be sufficiently strong to suppress corrections to the Liouville evolution, while remaining sufficiently weak that the classical Hamiltonian varies only slowly across an individual coarse-graining cell. 

For a general Hamiltonian we define the mismatch between the two Liouville generators by
\begin{equation}\label{eq:Cdef_main}
\mathcal C_{\boldsymbol\Delta} f
:=
L_{H_{\boldsymbol\Delta}}f-L_Hf
=
\{H_{\boldsymbol\Delta}-H,f\}.
\end{equation} 
For separable Hamiltonians $H=p^2/2m+V(x)$, the condition becomes especially transparent. Since Gaussian smoothing leaves the kinetic term unchanged, apart from an irrelevant additive constant, the discrepancy between the coarse-grained and original Liouville generators is entirely governed by the difference in the force terms:
\begin{equation}
    \mathcal C_{\boldsymbol\Delta}f
=
\bigl(V^{(1)}_{\boldsymbol\Delta}-V^{(1)}\bigr)\,\partial_p f.
\end{equation}
Moreover, on a bounded region $K\subset\mathbb R^2$,
\begin{equation}
\begin{aligned}
\|V^{(1)}_{\boldsymbol\Delta}-V^{(1)}\|_{L^\infty(K)}
&=
\left\|
\frac{\ell_x^2}{2}\int_0^1 e^{s\frac{\ell_x^2}{2}\partial_x^2}V^{(3)}\,ds
\right\|_{L^\infty(K)} \\
&\le
\frac{\ell_x^2}{2}\int_0^1
\|e^{s\frac{\ell_x^2}{2}\partial_x^2}V^{(3)}\|_{L^\infty(K)}\,ds \\
&\le
\frac{\ell_x^2}{2}\sup_{0\le s\le 1}
\|e^{s\frac{\ell_x^2}{2}\partial_x^2}V^{(3)}\|_{L^\infty(K)}.
\end{aligned}
\end{equation}
To interpret this bound, let $L_F(K)$ denote the characteristic length scale over which the derivative of the force varies across the region $K$, for instance, by defining 
\begin{equation}
L_F(K):=\frac{\|V^{(2)}\|_{L^\infty(K)}}{\|V^{(3)}\|_{L^\infty(K)}}.
\end{equation}
Then $V^{(3)}\sim V^{(2)}/L_F(K)$, and the force-mismatch estimate becomes
\begin{equation}
\|V^{(1)}_{\boldsymbol\Delta}-V^{(1)}\|_{L^\infty(K)}
\lesssim
\frac{\ell_x^2}{L_F(K)}\,\|V^{(2)}\|_{L^\infty(K)}.
\end{equation}
To make this smallness condition dimensionless, it is natural to compare the force mismatch to the typical variation of the force across one coarse-graining cell, namely
\[
\ell_x\,\|V^{(2)}\|_{L^\infty(K)}.
\]
Hence
\begin{equation}
\frac{\|V^{(1)}_{\boldsymbol\Delta}-V^{(1)}\|_{L^\infty(K)}}
{\ell_x\,\|V^{(2)}\|_{L^\infty(K)}}
\lesssim
\frac{\ell_x}{L_F(K)}.
\label{eq:force_relative}
\end{equation}
Thus the correction is relatively small whenever
\begin{equation}
    \frac{\ell_x}{L_F(K)} \ll 1.
    \label{force}
\end{equation}
In physical terms, this means that the derivative of the force is approximately linear across a single coarse-graining cell, so replacing it by its Gaussian average does not significantly modify the classical flow.

Importantly, condition~\eqref{force} is naturally satisfied in the \emph{macroscopic limit}, that is, in the regime where the characteristic length scale over which the derivative of the force varies is itself macroscopic, say of order $ \sim 10^{-4}\,\mathrm{m} - 1\,\mathrm{m} $. Indeed, even for extreme coarse graining of the position of a macroscopic body of mass $1\,\mathrm{kg}$, the condition in Eq.~\eqref{force} is exceedingly well satisfied. This can be seen from the fact that the thermal momentum is of order $p_{\mathrm{th}}\sim \sqrt{3mk_B T}$, so that the associated de Broglie wavelength is
\begin{equation}
\lambda_{\mathrm{dB}} \sim \frac{h}{\sqrt{3mk_B T}} \sim 10^{-24}\,\mathrm{m}
\end{equation}
for $m=1\,\mathrm{kg}$ and room temperature of $T\sim 300\,\mathrm{K}$. Thus, even under very strong coarse graining, \(l_x \gg \Lambda_{\mathrm{dB}}\), condition~\eqref{force} remains exceedingly well satisfied for regions over which the relevant derivatives change only on macroscopic length scales, say of order \(1\,\mathrm{mm}\)--\(1\,\mathrm{cm}\). This shows that, when \emph{coarse graining is combined with the macroscopic limit}, the smoothed Hamiltonian becomes effectively equivalent in functional form to the classical Hamiltonian whose quantization yields the underlying quantum Hamiltonian. This provides a formal expression of the classical--quantum--classical loop illustrated in Fig.~\ref{fig:1}.

\subsection{Example: Double-well potential}
\label{subsec:localized_double_well}
We now illustrate the general results with a simple non-quadratic example. The purpose of this example is to show how the abstract bounds derived above translate into transparent physical conditions for the emergence of classical dynamics in a concrete system.

We consider the quartic double-well Hamiltonian
\begin{equation}
\hat H=\frac{\hat p^2}{2m}+V(\hat x),
\qquad
V(x)=\lambda(x^2-a^2)^2.
\label{eq:dw_H}
\end{equation}
The potential has two minima at $x=\pm a$ and a barrier at $x=0$. If we focus on the left well and introduce the shifted coordinate
\begin{equation}
y:=x+a,
\end{equation}
the minimum is located at \(y=0\). Expanding the potential around \(x=-a\) then gives
\begin{equation}
V(-a+y)=4\lambda a^2y^2-4\lambda ay^3+\lambda y^4.
\label{eq:dw_local_expand}
\end{equation}
The quadratic part defines the local oscillation frequency
\begin{equation}
\omega:=\sqrt{\frac{V^{(2)}(-a)}{m}}
=\sqrt{\frac{8\lambda a^2}{m}},
\label{eq:dw_omega}
\end{equation}
so that, for $|y|\ll a$, the dynamics is approximately governed by the local harmonic Hamiltonian
\begin{equation}
H_{\rm loc}(y,p)=\frac{p^2}{2m}+\frac12 m\omega^2 y^2.
\label{eq:Hloc}
\end{equation}
The cubic and quartic terms in Eq.~\eqref{eq:dw_local_expand} quantify the anharmonic corrections and are small as long as the motion remains confined to a region much smaller than the well size $a$.

The full Hamiltonian~\eqref{eq:dw_H} provides one of the paradigmatic examples of a system exhibiting violations of the Leggett--Garg inequalities, and hence of macrorealism in the Leggett--Garg sense~\cite{leggett1985,EmaryLambertNori2014,Sakamoto2025LGI}. The mechanism responsible for this violation is quantum tunneling.

In the deep-well regime, where each localized packet is much narrower than the separation of the wells, the low-energy dynamics of a symmetric double well is well approximated by the two-dimensional subspace spanned by the lowest even $|+\rangle$ and odd $|-\rangle$ eigenstates~\cite{LeggettEtAl1987,Garg2000TunnelSplittings}. In this regime, the symmetric $|\psi_+\rangle$ and antisymmetric $|\psi_-\rangle$ Schrödinger cat states built from coherent packets  centered at the two minima approximate the lowest even and odd eigenstates of the double well, respectively:
\begin{equation}
|+\rangle \approx |\psi_{+}\rangle = \mathcal N\bigl(|-a,0\rangle + |a,0\rangle\bigr), \quad 
|-\rangle \approx |\psi_{-}\rangle = \mathcal N\bigl(|-a,0\rangle - |a,0\rangle\bigr),
\label{eq:cat_initial}
\end{equation}
where $\mathcal N$ is the normalization factor. 
Introducing the localized states
\begin{equation}
|R\rangle:=\frac{|+\rangle+|-\rangle}{\sqrt2} \approx |a,0\rangle,
\quad
|L\rangle:=\frac{|+\rangle-|-\rangle}{\sqrt2} \approx |-a,0\rangle,
\label{eq:LR_states}
\end{equation}
which represent, respectively, states localized predominantly in the right and left well, one obtains the standard effective two-level Hamiltonian on this subspace,
\begin{equation}
\hat H_{\rm eff}
=
\bar E\bigl(|L\rangle\langle L|+|R\rangle\langle R|\bigr)
-\frac{\Delta E}{2}\bigl(|L\rangle\langle R|+|R\rangle\langle L|\bigr),
\label{eq:Heff_dw}
\end{equation}
where $\bar E:=\frac{E_++E_-}{2}$ and $\Delta E:=E_--E_+$. Thus, if the particle is initially prepared in the right well, $|\psi(0)\rangle=|R\rangle$, then
\begin{equation}
P_R(t)=\cos^2\!\left(\frac{\Delta E\,t}{2\hbar}\right),
\qquad
P_L(t)=\sin^2\!\left(\frac{\Delta E\,t}{2\hbar}\right),
\label{eq:PLPR_dw}
\end{equation}
so that the first full transfer from one well to the other occurs at
\begin{equation}
t_{\rm tun}:=\frac{\pi\hbar}{\Delta E}.
\label{eq:t_tun_dw}
\end{equation}
In the deep-well regime, the splitting is exponentially small and, at leading semiclassical order, is controlled by the under-barrier action~\cite{Garg2000TunnelSplittings},
\begin{equation}
\Delta E \approx \frac{\hbar\omega}{\pi}\,e^{-S_0/\hbar},
\qquad
S_0=\int_{-a}^{a}dx\,\sqrt{2mV(x)}
=
\frac{4}{3}a^3\sqrt{2m\lambda}
=
\frac{2}{3}m\omega a^2.
\label{eq:DeltaE_WKB_dw}
\end{equation}
Hence
\begin{equation}
t_{\rm tun}\approx \frac{\pi^2}{\omega}\exp\!\left(\frac{S_0}{\hbar}\right)
=
\frac{\pi^2}{\omega}\exp\!\left(\frac{2m\omega a^2}{3\hbar}\right).
\label{eq:t_tun_WKB_dw}
\end{equation}
Even when the measurements are strongly coarse-grained so that they effectively distinguish only two broad spatial regions, to the left and to the right of the potential maximum, the violation of Leggett-Garg inequalities may still persist due to coherent interwell tunneling~\cite{leggett1985,EmaryLambertNori2014}. No classical Hamiltonian evolution can account for such oscillatory transfer.

The symmetric cat state~\eqref{eq:cat_initial}, as an approximation to $|+\rangle$, provides a useful example of how coarse graining suppresses distinctly quantum features at the kinematical level. In the deep-well approximation, this is an eigenstate of the Hamiltonian, and once the system is prepared in it, the state remains unchanged in time. For simplicity, we take the two coherent packets of the cat state to have the same widths $\sigma_x,\sigma_p$ as the coherent states entering the POVM construction. When $a\gg \sigma_x$, the overlap between the two branches is exponentially small, and the Wigner function is, up to exponentially small corrections,
\begin{equation}
W_0(x,p)
\approx
\frac12 W_{L}(x,p)+\frac12 W_{R}(x,p)+W_{\mathrm{int}}(x,p),
\label{eq:cat_W_split}
\end{equation}
with
\begin{equation}
W_{L/R}(x,p)
=
\frac{1}{\pi\hbar}
\exp\!\left(
-\frac{(x\pm a)^2}{2\sigma_x^2}
-\frac{p^2}{2\sigma_p^2}
\right)
\label{eq:cat_W_lobes}
\end{equation}
for the two positive Gaussian lobes, and
\begin{equation}
W_{\mathrm{int}}(x,p)
=
\frac{1}{\pi\hbar}
\exp\!\left(
-\frac{x^2}{2\sigma_x^2}
-\frac{p^2}{2\sigma_p^2}
\right)
\cos\!\left(\frac{2ap}{\hbar}\right)
\label{eq:cat_W_interf}
\end{equation}
for the interference term. The oscillatory factor $\cos(2ap/\hbar)$ produces fringes with characteristic momentum scale
\begin{equation}
\delta p_{\mathrm{int}}\sim \frac{\hbar}{a}.
\label{eq:fringe_scale}
\end{equation}
Momentum coarse graining suppresses the interference term by the factor
\begin{equation}
\eta(\ell_p):=
\exp\!\left(
-\frac{2a^2\sigma_p^2\ell_p^2}{\hbar^2(\sigma_p^2+\ell_p^2)}
\right),
\end{equation}
while also broadening its Gaussian envelope (and slightly renormalizing the fringe wave number). For our purposes, the key point is the amplitude suppression encoded by $\eta(\ell_p)$. In the strong coarse-graining regime $\ell_p\gg \sigma_p$, this becomes
\begin{equation}
\eta(\ell_p)\approx
\exp\!\left(-\frac{2a^2\sigma_p^2}{\hbar^2}\right)
=
\exp\!\left(-\frac{a^2}{2\sigma_x^2}\right),
\end{equation}
where we used $\sigma_x\sigma_p=\hbar/2$. Hence, in the deep-well regime $a\gg \sigma_x$, the interference contribution is exponentially suppressed. Therefore
\begin{equation}\label{eq:mixture_limit}
p_{\boldsymbol\Delta}(x,p,0)
\approx
\frac12 p_{L,\boldsymbol\Delta}(x,p,0)
+
\frac12 p_{R,\boldsymbol\Delta}(x,p,0),
\end{equation}
up to exponentially small corrections. Thus coarse graining suppresses the Wigner negativity/interference and renders the effective state operationally indistinguishable from a classical mixture of two localized packets.

\begin{figure}
\centering
\includegraphics[width=\linewidth]{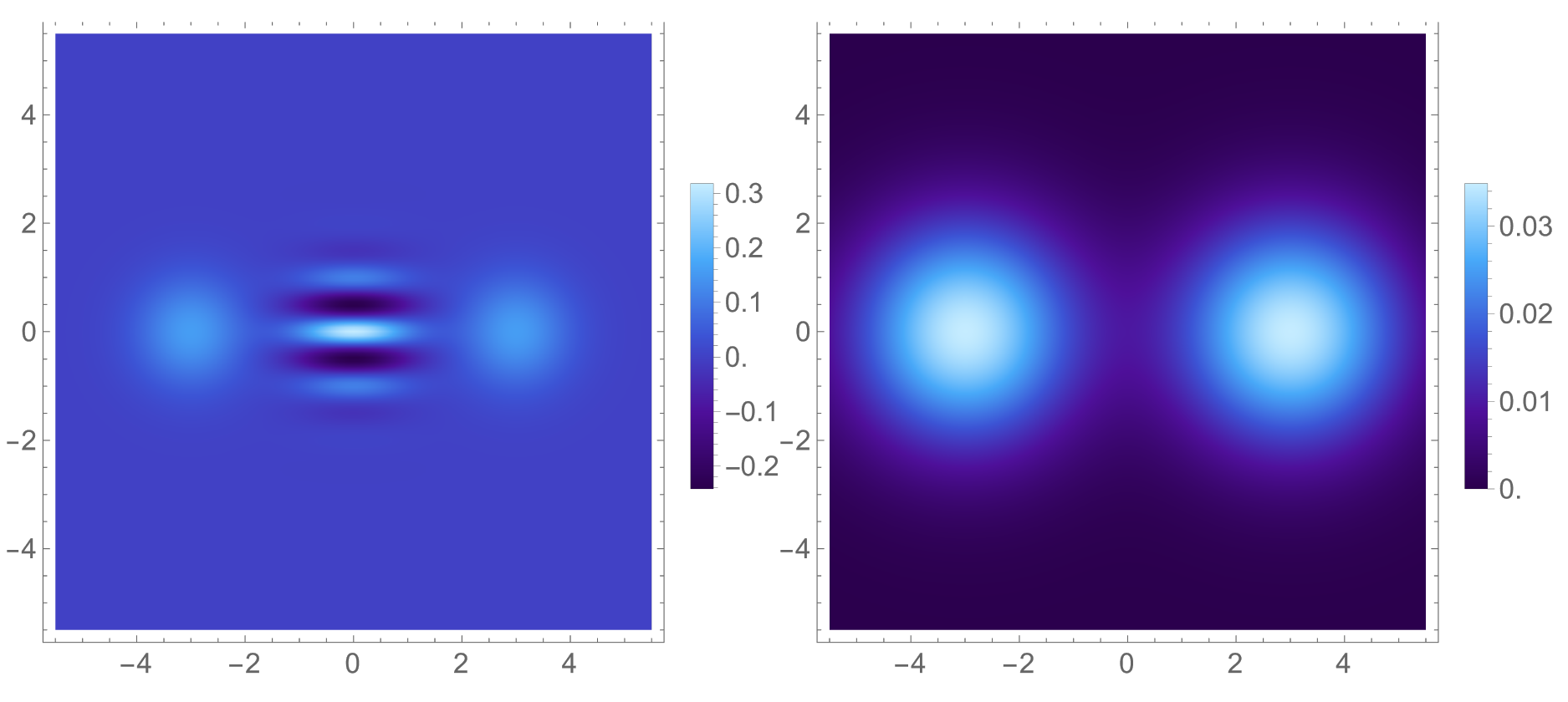}
\caption{On the left, the cat-state Wigner function of Eq.~\eqref{eq:cat_W_split}. On the right, its coarse-grained version $p_{\boldsymbol \Delta}=S_{\boldsymbol \Delta}W$. The interference term visible in the Wigner function disappears in the coarse-graining, yielding a fully classical probability distribution, undistinguishable from Eq.~\eqref{eq:mixture_limit}. Parameters used: $\hbar=1,\sigma_x=\sigma_p=\sqrt{1/2},a=3,\Delta_x=\Delta_p=1.6$.}
\label{fig:WvsP}
\end{figure}

Therefore, although the underlying quantum state is pure and coherent, the effective state induced by the coarse-grained POVMs becomes a classical mixture of two localized probability distributions. This shows how one of the paradigmatic examples of a nonclassical state appears effectively classical under coarse-grained measurements (see Fig.~\ref{fig:WvsP}). 

We now turn to investigating the dynamics generated by a double-well potential under coarse-grained measurements. From this point on, we restrict attention to the dynamics within one localized well, on times for which interwell leakage remains negligible at the resolution set by the POVM. This is important because the exact Schrödinger evolution of the full double well does exhibit coherent tunneling on the timescale $t_{\rm tun}$ derived above, whereas the estimates below concern the local coarse-grained dynamics within a single well. 

To render transparent the regime in which classicality emerges in the dynamics, it is convenient to introduce three dimensionless parameters.
\begin{equation}
\epsilon:=\frac{\ell_x}{a},
\qquad
r:=\frac{m\omega\,\ell_x}{\ell_p},
\qquad
q:=\frac{\hbar}{a\,\ell_p}.
\label{eq:dw_dimensionless}
\end{equation}
They have a simple interpretation. (i) The parameter $\epsilon$ compares the spatial coarse-graining scale with the size of one well. Thus $\epsilon\ll1$ means that one coarse-graining cell probes only the local geometry of a single well and does not resolve the global double-well structure. (ii) The parameter $r$ compares the momentum resolution $\ell_p$ with the classical momentum variation $m\omega\ell_x$ generated by the local harmonic flow across one spatial cell. Hence $r\lesssim1$ means that the momentum coarse graining is at least as large as the local classical momentum variation. (iii) Finally, $q$ compares the interference scale $\hbar/a$ in momentum with the coarse-graining width $\ell_p$; therefore $q\ll1$ means that momentum coarse graining washes out interference fringes associated with coherence over a distance of order $a$.

We now specialize the general bounds of Statements~\ref{st:D_bound} and~\ref{st:R_bound_poly} to the present example. For the quartic double well,
\begin{equation}
\begin{aligned}
V^{(1)}(x) &= 4\lambda x(x^2-a^2), \qquad
V^{(2)}(x) = 12\lambda x^2-4\lambda a^2, \\
V^{(3)}(x) &= 24\lambda x, \qquad \qquad  \quad \ \
V^{(4)}(x) = 24\lambda .
\end{aligned}
\label{eq:dw_derivatives}
\end{equation}
while $V^{(k)}\equiv0$ for all $k\ge5$. We consider times such that the evolved localized quantum state remains within a one-well region $K_t$, centered around the local minimum and of size $O(\ell_x)$. In this regime, since $\ell_x\ll a$, one has the local estimates
\begin{equation}
V^{(2)}(x)\sim m\omega^2,
\qquad
V^{(3)}(x)\sim \frac{m\omega^2}{a},
\qquad
V^{(4)}(x)\sim \frac{m\omega^2}{a^2},
\label{eq:dw_local_scales}
\end{equation}
up to numerical constants uniform on $K_t$.

The classical commutator correction $\mathcal D_{\boldsymbol\Delta}$ is then bounded by
\begin{equation}
\begin{aligned}
\Gamma_{\mathcal D}(t)
&:=
\sup_{0\le s\le t}
\|\mathcal D_{\boldsymbol\Delta}p_{\boldsymbol\Delta}(s)\|_{L^\infty(K_{t-s})}
\\
&\lesssim
\frac{1}{2\pi}
\sup_{0\le s\le t}\|W(s)\|_{L^1(\mathbb R^2)}
\left[
\frac{1}{m\ell_x^2}
+
\frac{m\omega^2}{\ell_p^2}
\bigl(1+c_1\epsilon+c_2\epsilon^2\bigr)
\right].
\end{aligned}
\label{eq:dw_GammaD_clean}
\end{equation}
for universal constants $c_1,c_2>0$. Rewriting the second term in terms of $r$ gives the more transparent form
\begin{equation}
\Gamma_{\mathcal D}(t)
\;\lesssim\;
\frac{1}{2\pi}
\sup_{0\le s\le t}\|W(s)\|_{L^1(\mathbb R^2)}
\,
\frac{1}{m\ell_x^2}
\left[
1+r^2\bigl(1+c_1\epsilon+c_2\epsilon^2\bigr)
\right].
\label{eq:dw_GammaD_r}
\end{equation}
This shows that $\mathcal D_{\boldsymbol\Delta}$ is small provided $\epsilon \ll1,$ and $
r\lesssim 1.$
The first condition ensures that the force is approximately linear across one coarse-graining cell, while the second ensures that the local phase-space shear generated by the classical flow is not resolved by the momentum coarse graining.

The genuinely quantum correction $\mathcal R_{\boldsymbol\Delta}$ is even more strongly suppressed. Since the quartic potential satisfies $V^{(5)}\equiv0$, the Moyal remainder truncates after the first nontrivial term, and Statement~\ref{st:R_bound_poly} yields
\begin{equation}
\Gamma_{\mathcal R}(t)
:=
\sup_{0\le s\le t}
\|\mathcal R_{\boldsymbol\Delta}p_{\boldsymbol\Delta}(s)\|_{L^\infty(K_{t-s})}
\;\lesssim\;
\frac{\lambda\hbar^2}{\ell_x\ell_p^4}
\sup_{0\le s\le t}\|x\,W(s)\|_{L^1(\mathbb R^2)}.
\label{eq:dw_GammaR_clean}
\end{equation}
Inside one well, $|x|\sim a$, so schematically
\begin{equation}
\Gamma_{\mathcal R}(t)
\;\lesssim\;
\frac{m\omega^2\,\hbar^2}{a\,\ell_x\,\ell_p^4}
\sup_{0\le s\le t}\|W(s)\|_{L^1(\mathbb R^2)},
\label{eq:dw_GammaR_clean2}
\end{equation}
again up to a numerical constant. The decisive feature is the strong $\ell_p^{-4}$ suppression. In particular, the condition
$q=\hbar/a\ell_p\ll1$ ensures that the coarse graining does not resolve quantum structure on the momentum scale $\hbar/a$.

\begin{figure}
\centering
\includegraphics[width=1\linewidth]{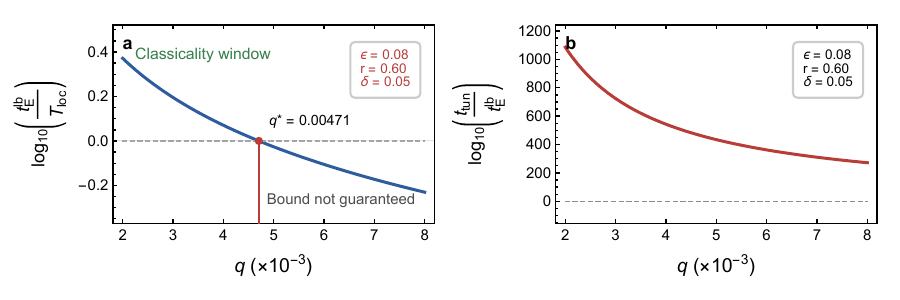}
\caption{
Crossover in the existence of a coarse-grained classicality window for the double-well example at fixed $\epsilon$, $r$ and $\delta$. The plotted quantity in panel $a$ is $\log_{10}(t_E^{\mathrm{lb}}/T_{\mathrm{loc}})$, where $T_{\mathrm{loc}}=2\pi/\omega$ is the local oscillation period in one well. The critical value $q^*$ is defined by $t_E^{\mathrm{lb}}(q^*)=T_{\mathrm{loc}}$, where $q=\hbar/a\,\ell_p$. For $q<q^*$, the Ehrenfest lower bound exceeds one local oscillation period, so the bound guarantees a classicality window. For $q>q^*$, this lower bound no longer guarantees such a window. In the same parameter range as depicted in panel $b$ one still has $t_{\mathrm{tun}}\gg t_E^{\mathrm{lb}}$, so tunneling remains parametrically later.  Here we set parameters $\hbar=m=\omega=1$.
}
\label{fig:dw_classical_window}
\end{figure}

Combining the above estimates with the Duhamel bound of Statement~\ref{st:duhamel}, we conclude that the coarse-grained dynamics is well approximated by Liouville transport whenever $
\epsilon \ll1,$ $r\lesssim1,$ and $
q\ll1.$ In this regime, the coarse-grained Ehrenfest time is correspondingly large. Indeed, from Eq.~\eqref{eq:tE-quant-estimate},
\begin{equation}
t_E
\gtrsim
\frac{\delta}{\Gamma_{\mathcal D}(0)+\Gamma_{\mathcal R}(0)},
\label{eq:dw_tE}
\end{equation}
so that small correction rates imply a long time interval on which the effective phase-space density evolves essentially classically. Finally, within the one-well regime considered here, the smoothed Hamiltonian $H_{\boldsymbol\Delta}$ is itself close to the local harmonic Hamiltonian $H_{\rm loc}$.
To make this statement more concrete, it is useful to compare the Ehrenfest lower bound with the intrinsic local timescale of the one-well motion,
\begin{equation}
T_{\mathrm{loc}}:=\frac{2\pi}{\omega}.
\end{equation}
We say that a \emph{classicality window} is present whenever
\begin{equation}
T_{\mathrm{loc}}< t_E^{\mathrm{lb}} \ll t_{\mathrm{tun}},
\end{equation}
where $t_E^{\mathrm{lb}}$ denotes the lower bound in Eq.~\eqref{eq:dw_tE}. The first inequality means that the coarse-grained Liouville description is guaranteed to remain valid for at least one local oscillation period inside a single well, while the second ensures that coherent interwell tunneling sets in only on a much longer timescale. In this regime there exists an intermediate interval of times in which the measured dynamics is effectively classical, even though the exact quantum evolution still contains tunneling at exponentially later times. Figure~\ref{fig:dw_classical_window} illustrates this crossover at fixed $\epsilon$ and $r$ by plotting $\log_{10}(t_E^{\mathrm{lb}}/T_{\mathrm{loc}})$ as a function of $q$. The zero crossing, defined by $t_E^{\mathrm{lb}}=T_{\mathrm{loc}}$, marks the onset of the coarse-grained classicality window.

It is important to distinguish this coarse-grained Ehrenfest time from the tunneling time $t_{\rm tun}$. To compare the two, one should consider the same initial situation, namely a state initially localized in one well. The tunneling time then characterizes the exact quantum transfer to the opposite well, whereas the Ehrenfest time characterizes how long the corresponding coarse-grained dynamics within one well remains well approximated by local Liouville transport. The tunneling time grows exponentially when the wells are far apart compared with the local wavepacket width, or equivalently when the barrier is high compared with the local zero-point energy, cf.\ Eq.~\eqref{eq:t_tun_WKB_dw}. By contrast, the bound in Eq.~\eqref{eq:dw_tE} depends only algebraically on the coarse-graining parameters. Thus, in the deep-well regime, one typically expects a broad intermediate time window in which the coarse-grained dynamics within a single well is accurately described by local Liouville transport, even though the exact quantum dynamics still exhibits coherent interwell tunneling on much longer timescales and may therefore eventually lead to violations of the LG inequalities. If one wishes to suppress this tunneling physically, one needs coarse-grained measurements to be performed with a characteristic rate at least comparable to the tunneling frequency $\Delta E/\hbar\sim 1/t_{\rm tun}$. Since such monitoring acts on timescales much longer than the local Ehrenfest time, one would already expect deviations from the ideal one-well classical Liouville evolution before an actual interwell transfer becomes resolved.

\section{Emergence of classical trajectories}\label{sec:Trajectories}

In the previous section, we showed that, within the coarse-grained classicality window, the phase-space distribution $p_{\boldsymbol{\Delta}}(t)$ evolves approximately according to Liouville transport under the smoothed Hamiltonian
$H_{\boldsymbol{\Delta}} = S_{\boldsymbol{\Delta}} H$.
Operationally, this statement can be tested by performing measurements at various times $t_1,t_2,\dots,t_n$ on a large ensemble of identically prepared systems, with the protocol arranged so that, for the statistics collected at time $t_i$, no measurement is performed on a given system before $t_i$. This yields an ensemble-level notion of classicality. We now turn to an operational notion based on \emph{time-resolved measurement records} of a single system. The question is whether a sequence of finite-resolution phase-space
readouts can, with high probability, be regarded as following a classical trajectory.

The notion of trajectory formation considered here is of statistical nature. After observing a readout
$z_k$ at time $t_k$, a classical description is meaningful if the next readout $z_{k+1}$ at time
$t_{k+1}=t_k+\tau$ is likely to lie near the classically drifted point $\Phi_\tau(z_k)$, where $\Phi_\tau$ is the
Hamiltonian flow generated by $H_{\boldsymbol\Delta}$. At finite resolution, deviations from this ideal behaviour arise
for two distinct reasons: the measurement itself leaves an irreducible spread in phase space, and the exact
coarse-grained evolution over one sampling interval need not coincide exactly with Liouville transport. The discussion
below separates these two effects.

For concreteness, consider repeated measurements at times $t_k=k\tau$ with the Gaussian phase-space POVM
$\{\hat P_{\boldsymbol\Delta}(z)\}_{z\in\mathbb R^2}$, implemented by the L\"uders instrument
\begin{equation}\label{eq:traj_luders}
\hat K_{\boldsymbol\Delta}(z):=\sqrt{\hat P_{\boldsymbol\Delta}(z)},
\qquad
\hat P_{\boldsymbol\Delta}(z)=\hat K_{\boldsymbol\Delta}(z)^\dagger \hat K_{\boldsymbol\Delta}(z),
\end{equation}
with unitary evolution $\hat U(\tau)=e^{-i\hat H_{\boldsymbol\Delta}\tau/\hbar}$ between measurements. The post-measurement state
conditioned on outcome $z_k$ at time $t_k$ is
\begin{equation}\label{eq:traj_post_state}
\hat{\tilde\rho}_{z_k}
=
\frac{\hat K_{z_k}\hat \rho \hat K_{z_k}^\dagger}{p(z_k)},
\qquad
p(z_k):=\Tr[\hat P_{z_k}\hat \rho].
\end{equation}
The conditional density for the next outcome is obtained by evolving for a time $\tau$ and measuring again
\begin{equation}\label{eq:traj_conditional_exact}
p(z_{k+1}\mid z_k)
=
\Tr\!\left[\hat P_{z_{k+1}}\,\hat U(\tau)\hat{\tilde\rho}_{z_k}\hat U(\tau)^\dagger\right].
\end{equation}

As a classical reference, we consider the conditional density obtained by transporting the posterior distribution with
the classical flow. This gives
\begin{equation}\label{eq:traj_conditional_classical}
p^{\mathrm{cl}}(z_{k+1}\mid z_k)
:=
\Tr\!\left[\hat P_{\Phi_{-\tau}(z_{k+1})}\,\hat{\tilde\rho}_{z_k}\right].
\end{equation}
The result of the previous section implies that, on any bounded observation region $K\subset\mathbb R^2$,
\begin{equation}\label{eq:traj_dynamic_error}
\|p(\,\cdot\,|z_k)-p^{\mathrm{cl}}(\,\cdot\,|z_k)\|_{L^\infty(K)}
\le \delta_\tau,
\end{equation}
where $\delta_\tau$ is the one-step dynamical error. In particular, for $\tau=t_E$ we may write
$\delta_\tau=\delta$, where $\delta$ is the maximum tolerance introduced in Eq.~\eqref{eq:tE-quant-def}. This is the
dynamical contribution to the trajectory error.

The kinematical contribution comes from the finite width of the measurement itself. Even if the evolution were exactly
Liouville, the next outcome would not be sharply localized at the single point $\Phi_\tau(z_k)$, but only
concentrated around it with a finite spread set by the measurement resolution. Using Eq.~\eqref{eq:traj_post_state} in Eq.~\eqref{eq:traj_conditional_classical}, together with the operator inequality $0\le \hat\rho\le \mathbb{1}$, we obtain
\begin{equation}\label{eq:traj_pointwise_overlap_bound}
p^{\mathrm{cl}}(z_{k+1}\mid z_k)
\le
\frac{1}{p(z_k)}
\Tr\!\left[\hat P_{\Phi_{-\tau}(z_{k+1})}\hat P_{z_k}\right].
\end{equation}
For the Gaussian POVM, the overlap of two effects is explicitly
\begin{equation}\label{eq:traj_povm_overlap}
\Tr[\hat P_u\hat P_v]
=
\frac{1}{8\pi^2\hbar\,l_xl_p}
\exp\!\left[
-\frac{(u_x-v_x)^2}{4l_x^2}
-\frac{(u_p-v_p)^2}{4l_p^2}
\right].
\end{equation}
Thus the classical reference density is concentrated near those points $z_{k+1}$ whose backward image
$\Phi_{-\tau}(z_{k+1})$ lies close to the observed point $z_k$.

To quantify this spread, we introduce the resolution cell
\begin{equation}\label{eq:traj_resolution_cell}
E(z_k,\kappa)
:=
\left\{
u\in\mathbb R^2:
\frac{(u_x-x_k)^2}{4l_x^2}
+
\frac{(u_p-p_k)^2}{4l_p^2}
\le \kappa^2
\right\},
\end{equation}
which is the phase-space region around $z_k$ resolved by the measurement at tolerance $\kappa$. A natural notion of one-step classicality is then that the next readout lies inside the transported cell
$\Phi_\tau(E(z_k,\kappa))$. Using Eqs.~\eqref{eq:traj_pointwise_overlap_bound} and~\eqref{eq:traj_povm_overlap}, and
integrating over the complementary region, one obtains the bound
\begin{equation}\label{eq:traj_kinematical_tail}
p^{\mathrm{cl}}\!\left(
z_{k+1}\notin \Phi_\tau(E(z_k,\kappa))
\,\middle|\, z_k
\right)
=
\int_{\mathbb{R}^2\setminus \Phi_\tau(E(z_k,\kappa))} d^2u \, p^\mathrm{cl}(u\mid z_k)
\le
\frac{1}{2\pi\hbar\,p(z_k)}\,e^{-\kappa^2}.
\end{equation}
Thus, even under exact Liouville transport, finite measurement resolution leaves a nonzero probability of missing the
classically transported cell, but this probability is exponentially suppressed in $\kappa$.

Combining the kinematical estimate \eqref{eq:traj_kinematical_tail} with the dynamical bound
\eqref{eq:traj_dynamic_error}, we obtain a bound for the original probability as 
\begin{equation}\label{eq:traj_single_step_final}
\Pr\!\left(
z_{k+1}\notin \Phi_\tau(E(z_k,\kappa))
\,\middle|\, z_k
\right)
\le
\frac{1}{2\pi\hbar\,p(z_k)}\,e^{-\kappa^2}
+
|K|\,\delta_\tau .
\end{equation}
This is the basic one-step trajectory estimate. The first term is kinematical and reflects the finite width of the
measurement, while the second is dynamical and quantifies the deviation of the exact coarse-grained evolution from
Liouville transport over one sampling interval. Inside the classicality window, both are small, so the next readout
lies near the classically drifted point with high probability.

The same reasoning extends to longer records. Let
\begin{equation}\label{eq:traj_nstep_event}
\mathcal T_n(\kappa)
:=
\bigcap_{j=0}^{n-1}
\left\{
z_{j+1}\in \Phi_\tau(E(z_j,\kappa))
\right\},
\end{equation}
denote the event that the first $n$ readouts all remain inside the corresponding classically transported resolution
cells. If $p(z)\ge p_{\min}>0$ on the observation region $K$, then the one-step bound
\eqref{eq:traj_single_step_final} becomes uniform:
\begin{equation}\label{eq:traj_uniform_one_step_error}
\varepsilon_{\mathrm{step}}(\kappa)
:=
\frac{1}{2\pi\hbar\,p_{\min}}\,e^{-\kappa^2}
+
|K|\,\delta_\tau .
\end{equation}
A union bound then gives
\begin{equation}\label{eq:traj_nstep_bound}
\Pr\!\bigl(\mathcal T_n(\kappa)^c\bigr)\le n\,\varepsilon_{\mathrm{step}}(\kappa),
\qquad
\Pr\!\bigl(\mathcal T_n(\kappa)\bigr)\ge 1-n\,\varepsilon_{\mathrm{step}}(\kappa).
\end{equation}
Hence the probability of leaving the classical tube grows at most linearly with the number of sampling steps, while
each one-step error is controlled by the same two mechanisms as before: the Gaussian tails set by finite resolution
and the non-Liouville dynamical correction. In this operational sense, repeated coarse-grained phase-space measurements
generate, with high probability, a measurement record that remains confined to a tube around the classical trajectory. 

\begin{figure}[t]
    \centering

    \begin{subfigure}[t]{0.32\linewidth}
        \centering
        \includegraphics[width=\linewidth]{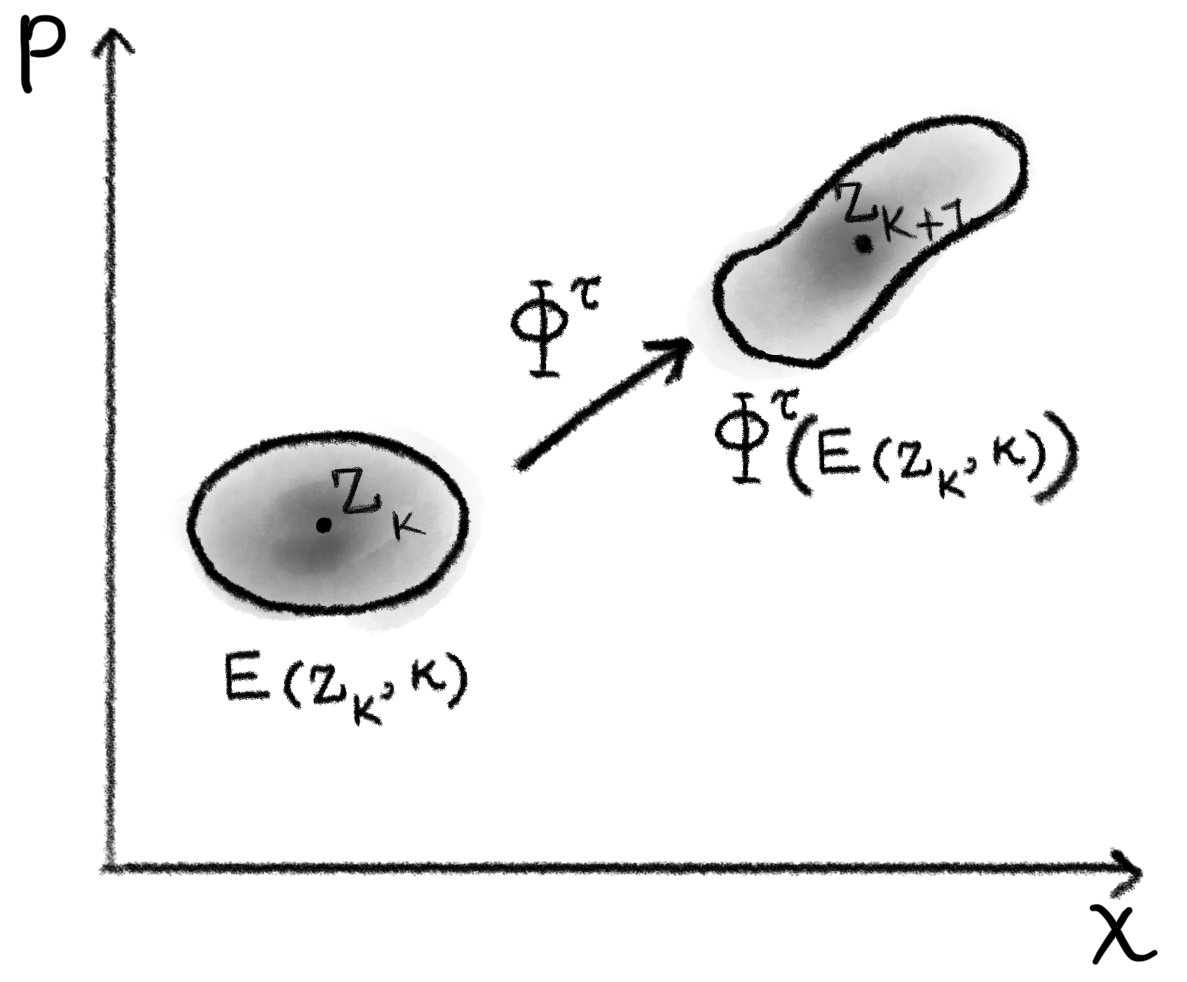}
        \caption{}
        \label{fig:Traj-schematic-a}
    \end{subfigure}
    \hfill
    \begin{subfigure}[t]{0.32\linewidth}
        \centering
        \includegraphics[width=\linewidth]{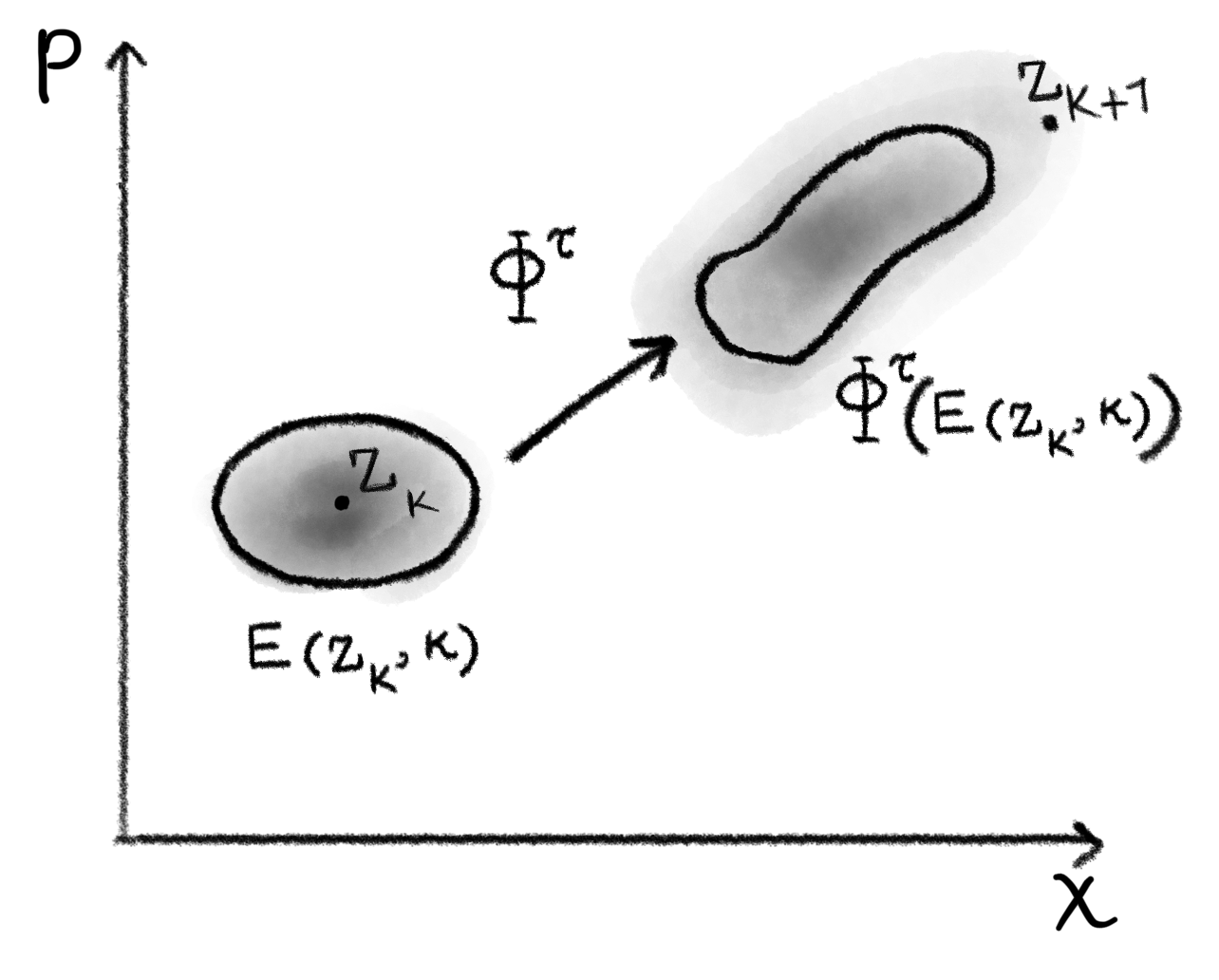}
        \caption{}
        \label{fig:Traj-schematic-b}
    \end{subfigure}
    \hfill
    \begin{subfigure}[t]{0.32\linewidth}
        \centering
        \includegraphics[width=\linewidth]{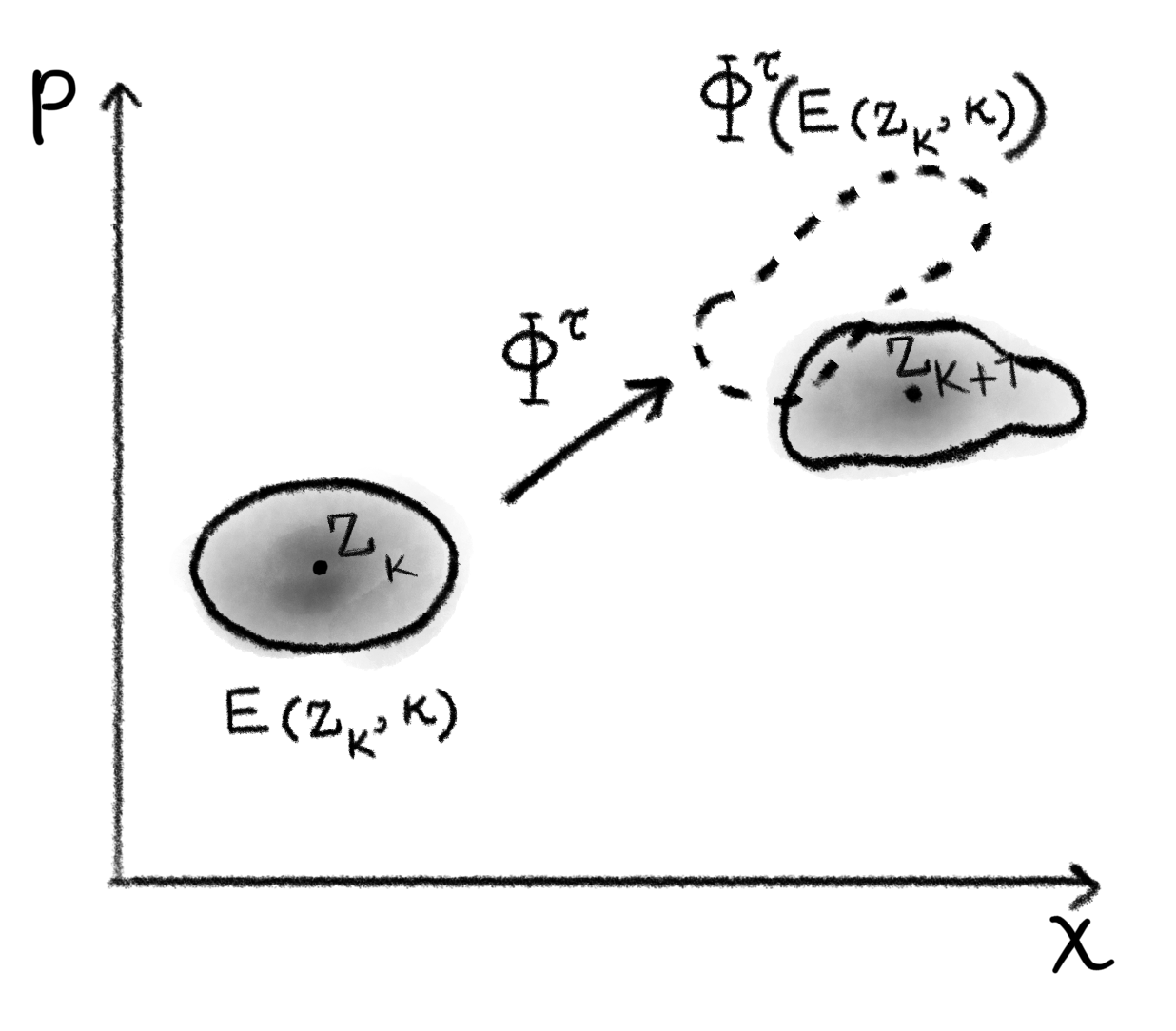}
        \caption{}
        \label{fig:Traj-schematic-c}
    \end{subfigure}

    \caption{
    Emergence of trajectory under repeated coarse-grained phase-space measurements. 
    Following a readout $z_k$, the relevant classical object is the transported resolution cell $\Phi_\tau(E(z_k,\kappa))$, namely the image of the full cell $E(z_k,\kappa)$ under the classical flow. 
    In general this image is not a rigidly shifted ellipse, but may be stretched and deformed by the flow. 
    (a) Ideal classical step: the next readout $z_{k+1}$ lies inside the transported cell. 
    (b) Kinematical error contribution: even under exact Liouville transport, the finite width of the Gaussian POVM leaves exponentially small tails outside $\Phi_\tau(E(z_k,\kappa))$. 
    (c) Dynamical error  contribution: the exact one-step predictive density need not coincide exactly with its Liouville-transported reference, producing an additional mismatch quantified by $\delta_\tau$. 
    Inside the classicality window, both effects are small, so the measurement record remains confined with high probability to a tube around the classical trajectory.
    }
\label{fig:traj_schematic}
\end{figure}

Strictly speaking, the trajectory notion established here is not that of a single predetermined classical orbit. Rather, each measurement outcome $z_k$ defines the center of the next classically expected resolution cell $\Phi_\tau(E(z_k,\kappa))$, and the measurement record is shown to remain, with high probability, inside this sequence of transported cells. The result is therefore a stochastic, resolution-limited trajectory in phase space: a tube-following record rather than a sharp microscopic path, as illustrated in Fig.~\ref{fig:traj_schematic}. This should be contrasted with the usual unmeasured Ehrenfest picture, where one compares the motion of a distinguished quantity such as the wavepacket centroid to a single classical orbit. Our analysis does not rely on such a centroid description. Instead, it provides an operational notion of classical trajectory directly at the level of time-resolved measurement outcomes, with an error that accumulates at most linearly in the number of measurement steps.

When the system is viewed on macroscopic scales, the distinction between a narrow tube and a sharp orbit becomes negligible. After rescaling phase-space coordinates by macroscopic units much larger than $(\ell_x,\ell_p)$, the transverse width of the tube tends to zero in the rescaled variables. In this sense, the stochastic resolution-limited trajectory obtained from repeated coarse-grained measurements becomes indistinguishable from an ordinary classical phase-space trajectory. 

After the rescaling, the tube converges to a phase-space trajectory $(x(t),p(t))$. The functions $x(t)$ and $p(t)$ satisfy Hamilton's equations for the classical Hamiltonian $ H_{\boldsymbol\Delta}(x,p)=\frac{p^{2}}{2m}+V_{\boldsymbol\Delta}(x)$, that is,
\begin{equation}
\dot{x}(t)=\frac{p(t)}{m}, \qquad
\dot{p}(t)=-\partial_x V_{\boldsymbol\Delta}\!\bigl(x(t)\bigr).
\end{equation}
This also allows one to introduce a notion of the classical force
$F(x):=-\partial_x V_{\boldsymbol\Delta}(x)$. Combining the two equations gives Newton's second law,
\begin{equation}
m\ddot{x}(t)=\dot{p}(t)=-\partial_x V_{\boldsymbol\Delta}\!\bigl(x(t)\bigr)=F\bigl(x(t)\bigr),
\end{equation}
thus showing that the limiting dynamics is governed by the Newtonian laws of mechanics. Hence, in the classical limit emerging from coarse-grained measurements, the notion of force in Newtonian mechanics is recovered starting solely from quantum potentials.

It is worth stressing that this notion of trajectory differs from the one usually studied in continuous-measurement theory. In the latter, one takes a weak-measurement limit in which the sampling interval tends to zero while the measurement strength is scaled simultaneously, leading to a stochastic master equation or quantum filtering equation for the conditioned state and to a noisy continuous measurement current \cite{bhattacharya2000continuous,bhattacharya2003quantum,ghose2003recovering,ghose2004transition,habib2006emergence,jacobs2006straightforward,brun2000continuous,javanainen2013emergent}. By contrast, our construction keeps both the sampling interval $\tau$ and the instrumental resolution $(\ell_x,\ell_p)$ finite, and it yields a direct non-asymptotic bound on the probability that the discrete record exits a classically transported phase-space tube. In this sense, the present result is not another derivation of continuous quantum trajectories, but an operational classicality statement for finite-rate, finite-resolution phase-space measurements, directly tied to the coarse-grained Liouville bounds derived above. Sending $\tau\to0$ at fixed measurement strength would in general not reproduce the standard continuous-monitoring limit, but instead drives the system toward the familiar Zeno regime of very frequent observations. However, taking this limit together with a adequate simultaneous scaling of the measurement strength would also recover the limit of continuous measurements.

\section{The classical limit in the microscopic and macroscopic systems}

\label{sec:Parameters}

In this section, we will discuss what our findings on the conditions for a classical limit of quantum mechanics predict for the observed classical behavior of macroscopic systems under standard conditions, as well as microscopic systems under repeated measurements as in the example of particle trajectories in a cloud chamber. We have already considered the concrete example of a particle in a double-well potential. Here, we provide another example. A cloud chamber is a particle detection device that visualizes the paths of charged particles by condensing a supersaturated vapor, typically alcohol, around the regions where the particles are detected. There are several seminal works explaining the emergence of classical trajectories for particles in the cloud chamber, ranging from the historical paper by Mott~\cite{Mott1929} to more modern approaches based on decoherence~\cite{zurek1991,Giulini1996} and consistent histories~\cite{GellMann1993}. Here, we provide a related approach that specifically focuses on the precision and rate of coarse-grained measurements to explain the classicality of microscopic systems.

We will make a comparative analysis of the Ehrenfest time~(\ref{free}) for both macro- and microsystems. We begin the analysis of alpha particles in a cloud chamber with an estimated measurement precision of $\ell_x\,=\, 10^{-10}\text{m}$. This corresponds to the effective radius $r_{\text{eff}}$ for an alpha particle interacting with an alcohol  molecule in a cloud chamber. Ionization in the chamber occurs when the Coulomb potential energy between the alpha particle and a molecule is comparable to the molecule's ionization energy $E_{\text{ion}}$. The Coulomb potential is given by $
V(r) = \frac{q_\alpha q_{\text{mol}}}{4 \pi \epsilon_0 r}
$ with the charge $q_{\alpha}$ of the alpha particle and $q_{\text{mol}}$ of the molecule. Setting $V(r_{\text{eff}}) = E_{\text{ion}}$, the effective radius is:
$
r_{\text{eff}} = \frac{q_\alpha q_{\text{mol}}}{4 \pi \epsilon_0 E_{\text{ion}}}.
$ Substituting typical values ($q_\alpha = 2e$, $q_{\text{mol}} = e$, $E_{\text{ion}} \sim 10 \, \text{eV} )$, one obtains that the necessary localization of the particle to produce ionization must be of order of $
r_{\text{eff}} \equiv \ell_x \sim 10^{-10} \, \text{m}. $  

The typical velocities of alpha particles emitted from radioactive decay (e.g., $5 \, \text{MeV}$) are around $v \sim 10^7 \, \text{m/s}$. 
The precision $\ell_p$ in measuring (perpendicular) momentum can be obtained from the relation 
\begin{equation}
    \Delta p = 2p \sin\left(\frac{\Delta \theta}{2}\right),
\end{equation} 
where $p=mv \sim 10^{-19} \text{kg m/s}
$ is the initial (transversal) momentum of the alpha particle and $\Delta \theta$ is the scattering angle from a collision. A finite resolution in the measurement of the width $\Delta L$ of the alpha particle trace translates to a finite angular resolution:
\begin{equation}
\Delta \theta \sim \frac{\Delta L}{L},
\end{equation}
where $L$ is the distance between the scattering point and the detection spot. Typically, the ionization trail left by an alpha particle has a radius on the order of microns, $\Delta L \sim 10^{-6} \, \text{m}$. For $L \sim 10^{-1} \, \text{m}$, this gives $\Delta \theta \sim 10^{-5} \, \text{rad}.$ The resulting precision in momentum is $\ell_ p=10^{-24} \text{kg m/s}$.
We now estimate the corresponding Ehrenfest time for the free evolution between two successive ionization events. For a free particle, the quantum correction vanishes,
\begin{equation}
\Gamma_{\mathcal R}(t)=0,
\end{equation}
and the only contribution comes from the classical commutator term,
\begin{equation}
\Gamma_{\mathcal D}(t)
=
\sup_{0\le s\le t}
\frac{1}{2\pi}\,\|W(s)\|_{L^1(\mathbb R^2)}
\frac{e^{-1}}{m\ell_x^2}.
\label{eq:cloud_GammaD}
\end{equation}
Assuming that after the interaction with the vapor, the interference in the state of the particles is negligible on phase space, i.e. $\|W(s)\|_{L^1(\mathbb{R}^2)}\approx 1$, this gives
\begin{equation}
\Gamma_{\mathcal D}(t)\approx \frac{e^{-1}}{2\pi m\ell_x^2}.
\end{equation}

To turn this into a quantitative estimate for $t_E$, we choose the tolerance $\delta$ as a fixed fraction $\eta\in(0,1)$ of the peak height of the coarse-grained phase-space distribution. Since the latter is Gaussian with widths $\ell_x,\ell_p$, its maximum value is
\begin{equation}
p_{\max}=\frac{1}{2\pi \ell_x\ell_p}.
\end{equation}
Setting
\begin{equation}
\delta=\eta\,p_{\max}
=
\frac{\eta}{2\pi \ell_x\ell_p},
\label{eq:cloud_delta}
\end{equation}
the general estimate $t_E\gtrsim \delta/\Gamma_{\mathcal D}(0)$ yields
\begin{equation}
t_E
\gtrsim
\eta\,e\,\frac{m\ell_x}{\ell_p}.
\label{eq:cloud_tE_est}
\end{equation}

Substituting the values
\begin{equation}
m_\alpha \approx 6.6\times 10^{-27}\,\mathrm{kg},
\qquad
\ell_x\sim 10^{-10}\,\mathrm{m},
\qquad
\ell_p\sim 10^{-24}\,\mathrm{kg\,m/s},
\end{equation}
we obtain
\begin{equation}
t_E
\gtrsim
\eta\,e\,
\frac{(6.6\times10^{-27})(10^{-10})}{10^{-24}}
\;\mathrm{s}
\;\approx\;
\eta\,1.8\times 10^{-12}\,\mathrm{s}.
\label{eq:cloud_tE_num}
\end{equation}
For instance, for a $10\%$ tolerance ($\eta=0.1$), this gives
\begin{equation}
t_E \sim 10^{-13}\,\mathrm{s}.
\end{equation}
Taking the lower bound in Eq.~\eqref{eq:cloud_tE_num} as an estimate of the minimal Ehrenfest time, we conclude that it is extremely short for microscopic systems. In the cloud chamber this means that classical behavior can persist only if the alpha particle is re-localized sufficiently often. This is precisely what happens physically: each ionizing collision with a vapor molecule acts as a coarse-grained position measurement.

The time between two such collisions is determined by the mean free path $\lambda$ of the alpha particle and its velocity $v$,
\begin{equation}
\lambda=\frac{1}{n\sigma},
\qquad
\tau=\frac{\lambda}{v},
\end{equation}
where $n$ is the molecular number density in the chamber and $\sigma$ is the effective scattering cross section. For typical values
\begin{equation}
n\sim 10^{25}\,\mathrm{m}^{-3},
\qquad
\sigma\sim 10^{-18}\,\mathrm{m}^2,
\qquad
v\sim 10^7\,\mathrm{m/s},
\end{equation}
one finds
\begin{equation}
\lambda\sim 10^{-7}\,\mathrm{m},
\qquad
\tau\sim 10^{-14}\,\mathrm{s}.
\end{equation}
This is shorter than the estimate $t_E\sim 10^{-13}\,\mathrm{s}$ obtained above for a $10\%$ tolerance. Hence, before the free quantum evolution can deviate appreciably from the Liouville approximation, the next ionization event already occurs, thereby re-localizing the particle. In this sense, the observed particle trajectory in the cloud chamber emerges from a sequence of sufficiently frequent coarse-grained measurements, each of which restores the conditions under which the classical approximation remains valid until the next collision of the particle with the vapor molecules.

We now turn to a macroscopic object under ordinary laboratory conditions. Standard optical instruments resolve position only up to about
\begin{equation}
\ell_x\sim 10^{-6}\,\mathrm{m}.
\end{equation}
For an object of mass $m=1\,\mathrm g=10^{-3}\,\mathrm{kg}$ at room temperature $T\approx 300\,\mathrm K$, the thermal de Broglie wavelength is
\begin{equation}
\lambda_{\mathrm{DB}}
=
\frac{h}{\sqrt{3mk_BT}}
\sim 10^{-22}\,\mathrm{m},
\end{equation}
so that
\begin{equation}
\frac{\ell_x}{\lambda_{\mathrm{DB}}}\sim 10^{16}\gg 1.
\end{equation}
Thus even a very good laboratory measurement is enormously coarse compared with the intrinsic quantum spatial scale of the object.

As a natural coarse-graining scale in momentum, we take multiples of the thermal momentum spread
\begin{equation}
\ell_p \gtrsim p_{\mathrm{th}}:=\sqrt{3mk_BT}
\approx 3.5\times 10^{-12}\,\mathrm{kg\,m/s}.
\end{equation}
Then the coarse-grained phase-space cell satisfies
\begin{equation}
\frac{\ell_x\ell_p}{\hbar}
\gtrsim
\frac{(10^{-6})(3.5\times10^{-12})}{10^{-34}}
\sim 10^{16},
\end{equation}
which places the system deep in the strong coarse-graining regime of Statement~\ref{prop:Decay}. In particular, the corresponding coarse-grained observables are already overwhelmingly close to commuting.

Using again the free-particle estimate~\eqref{eq:cloud_tE_est}, we obtain
\begin{equation}
t_E
\gtrsim
\eta\,e\,\frac{m\ell_x}{\ell_p}
\approx
\eta\,e\,
\frac{(10^{-3})(10^{-6})}{3.5\times10^{-12}}
\,\mathrm s
\approx
\eta\,7.7\times 10^2\,\mathrm s.
\end{equation}
For a $10\%$ tolerance ($\eta=0.1$) this gives
\begin{equation}
t_E\sim 10^2\,\mathrm s.
\end{equation}
Taken by itself, this lower bound may not appear particularly large. However, the physically relevant comparison is not with human timescales, but with the timescale on which the environment continuously \enquote{measures} and re-localizes a macroscopic object. Under ordinary atmospheric conditions, the flux of air molecules is of order
\begin{equation}
\Phi \sim \frac14 n_{\mathrm{air}}v_{\mathrm{th}}
\simeq 3\times 10^{27}\,\mathrm{m^{-2}s^{-1}}.
\end{equation}
Here we used
\(n_{\mathrm{air}}\simeq P/(k_BT)\simeq 2.4\times10^{25}\,\mathrm{m^{-3}}\)
for \(P=1\,\mathrm{atm}\) and \(T\simeq300\,\mathrm K\), and
\(v_{\mathrm{th}}\simeq \sqrt{8k_BT/(\pi m_{\mathrm{air}})}
\simeq 4.7\times10^2\,\mathrm{m/s}\), with
\(m_{\mathrm{air}}\simeq 4.8\times10^{-26}\,\mathrm{kg}\). For a macroscopic body with exposed area $A\sim 10^{-4}\text{--}10^{-3}\,\mathrm m^2$, this corresponds to a collision rate
\begin{equation}
R\sim \Phi A \sim 10^{23}\text{--}10^{24}\,\mathrm s^{-1},
\end{equation}
that is, a time between successive environmental interactions of only
\begin{equation}
\tau_{\mathrm{env}}\sim 10^{-24}\text{--}10^{-23}\,\mathrm s.
\end{equation}
Thus even the conservative lower bound $t_E\sim 10^2\,\mathrm s$ exceeds the environmental re-localization time by roughly 25 orders of magnitude. In this sense, classicality of macroscopic bodies is overwhelmingly stable: the environment monitors the system far more rapidly than any noticeable deviation from Liouville evolution could accumulate.

The contrast is therefore clear. For microscopic particles, the Ehrenfest time associated with a given coarse-grained measurement is very short, so classical behavior requires repeated measurements at a rate faster than $t_E^{-1}$. In a cloud chamber this condition is met by the rapid sequence of ionizing collisions. For macroscopic bodies, by contrast, the observational coarse graining is already extremely large compared with the relevant quantum scales, and environmental monitoring occurs on timescales incomparably shorter than the corresponding Ehrenfest bound. Our results therefore predict that classical behavior of macroscopic systems under standard conditions is generic, whereas for microscopic systems it emerges only when sufficiently frequent coarse-grained measurements repeatedly enforce the classical regime.

We summarize the physical conditions for alpha particle and a macroscopic mass of $1g$ in the following table. 

\begin{equation}
\begin{array}{|l|c|c|}
\hline
\textbf{Property} & \textbf{Micro } & \textbf{Macro} \\
\hline
\text{Mass (kg)} & 10^{-27} & 10^{-3} \\
\text{Position measurement precision $l_x$ (m)} & 10^{-10} & 10^{-6} \\
\text{Momentum measurement precision $l_p$ (kg·m/s)} & 10^{-24} & 10^{-12} \\
\text{The Ehrenfest time $t_E$ (s)} & 10^{-13} & 10^{2} \\
\hline
\end{array}
\end{equation}

\section{Conclusion and outlook}\label{sec:Conclusion}

This study provides a comprehensive framework for understanding how classical mechanics emerges from quantum mechanics through coarse-grained measurements. We have shown that classical kinematics arises naturally when the phase-space resolution of the measurements is much coarser than the scale set by Planck's constant. Such coarse-grained measurements are described by POVM elements defined as Gaussian-weighted integrals of coherent states over phase-space regions whose area exceeds Planck's constant, the centers of these regions representing the associated classical measurement outcomes. The set of all such outcomes defines a secondary operational phase space, and the probability density induced on this space is positive. This establishes the kinematical classical limit under coarse-grained measurements.

Furthermore, we established that classical dynamics emerges naturally under coarse-grained measurements. We derived the evolution of the probabilities associated with the coarse-grained measurements and showed that it consists of three contributions: Liouville evolution generated by the Hamiltonian smoothed over the POVM region, diffusion terms arising solely from coarse-graining, and a genuinely quantum reminder due to the fact that phase-space dynamics in quantum mechanics is governed by the Moyal bracket, with its higher-derivative corrections. We showed, however, that for sufficiently large coarse-graining the evolution is effectively reduced to the Liouvillian contribution, at least up to the Ehrenfest time, defined as the maximal time over which the error made in replacing the full evolution by the classical Liouville evolution remains tolerable.

Our analysis of the Ehrenfest time revealed its critical role in defining the time scale within which classical approximations remain valid. For macroscopic systems, this time scale is exceedingly long, ensuring classical behavior under standard conditions. Conversely, for microscopic systems, the Ehrenfest time is short, and classicality can only be preserved through frequent measurements, such as those in cloud chambers. Under such frequent measurements, one can reconstruct a classical ``tube'' that approximately follows the classical Hamiltonian flow and that, after an appropriate rescaling (i.e. ``zooming out'' the picture), converges to a classical phase-space trajectory.

From a foundational point of view, our work shows that although quantum theory places fundamental limits on the classical notion of objectivity through contextuality and quantum nonlocality, it can, under coarse-grained measurements, effectively restore it, thereby giving rise to the form of objectivity familiar from everyday experience.

Our results have broader implications for the procedure of quantization. In the Dirac quantization procedure, one starts from a classical Hamiltonian and then applies quantization rules to obtain the quantum Hamiltonian. This can be considered as one path from classical to quantum physics. What our work shows is that, in the reverse direction, coarse-grained measurements together with the macroscopic limit---in which the scale over which the Hamiltonian varies appreciably is itself macroscopic---lead back to the same effective classical Hamiltonian from which the quantization procedure originally started. In this way, we demonstrate the consistency of the quantization-classical limit loop.

From a more general perspective, our approach belongs to the broader class of approaches to the classical limit within quantum mechanics. One common route to the classical limit is based on the assumption that Planck’s constant is much smaller than the action for certain trajectories, as emphasized in Feynman’s path integral formulation. In this view, the sum over paths is strongly peaked around classical trajectories. Another well-established approach is decoherence, where interactions with the environment suppress interference effects for quantum systems. However, both approaches still leave open the possibility of observing quantum effects in principle, provided that one employs a highly sophisticated and utterly impractical measurement device with sufficient precision to access the degrees of freedom of the large system and its environment. By explicitly taking into account the precision limits of measurement apparatuses, below which quantum effects can no longer be retrieved, the coarse-grained approach is best seen not as competing with, but rather as complementary to, existing approaches to the classical limit, whose primary focus is on epistemic limitations.

Our findings provide a solid foundation for a general approach to the quantum-to-classical transition, bridging theoretical and practical insights. However, many questions remain to be addressed. Among them is how the present derivation of the classical limit of quantum mechanics in non-relativistic domain can be extended to the relativistic and field-theoretical domain. Can we, in a similar vein, obtain classical field equations from quantum field analysis in the limit of coarse-grained measurements? The initial analysis of this problem can be found in Ref.~\cite{Costa2013}. What are the appropriate coarse-grained POVMs in the quantum relativistic domain that would enable us to derive classical relativistic mechanics from quantum field theory and quantum relativistic physics?

Such advancements hold promise for deepening our understanding of the quantum-to-classical transition and could provide new tools for probing macroscopic quantum phenomena, both experimentally and theoretically.

\section*{Acknowledgements}
This research was funded in whole or in part by
the Austrian Science Fund (FWF) [10.55776/F71] and
[10.55776/COE1]. This publication was made possible through the
financial support of WOST (WithOutSpaceTime) grant
from the John Templeton Foundation. The opinions expressed in this publication are those of the authors and
do not necessarily reflect the views of the John Templeton Foundation
For the purpose of open access, the author(s) have applied a CC BY license to any author-accepted manuscript version arising from this submission.

\appendix
\section{Proofs of the statements} \label{app:proofs}
\subsection{Statement 1} \label{app:statement1}
For simplicity write the coarse-grained observable of Eq.~\eqref{eq:CGo} as
\begin{equation}
\hat A_{\boldsymbol\Delta}
=\frac{1}{2\pi\hbar}\int_{\mathbb R^2} A_{\boldsymbol\Delta}(z)\,|z\rangle\langle z|\,dz,
\qquad z=(x,p).
\end{equation}
We will rewrite $\hat A_{\boldsymbol\Delta}$ in Weyl form, express the commutator through Weyl operators, and finally bound the resulting integral.

We now introduce the Weyl inversion formula for coherent-state projectors, using the symplectic matrix $J=\begin{pmatrix}0&1\\-1&0\end{pmatrix}$ and the standard deviation matrix $\Sigma=\mathrm{diag}(\sigma_x,\sigma_p)$. The inversion formula reads
\begin{equation}
|z\rangle\langle z|
=\frac{1}{2\pi\hbar}\int_{\mathbb R^2}
e^{-\frac{i}{\hbar}\,z^{\mathsf T}J\xi}\;
e^{-\frac14\,\xi^{\mathsf T}\Sigma^{-2}\xi}\;
D(\xi)\,d\xi,
\end{equation}
 where $D(\xi)=e^{\frac{i}{\hbar} \hat z^{\mathsf T} J \xi}$ is the displacement operator (with $\hat z=(\hat x,\hat p)$). We replace the Weyl inversion formula in the coarse-grained observable expression, and replace the symplectic Fourier transform
\begin{equation}
\widetilde A_{\boldsymbol\Delta}(\xi)
=\frac{1}{2\pi\hbar}\int_{\mathbb R^2}
A_{\boldsymbol\Delta}(z)\,e^{-\frac{i}{\hbar}\,z^{\mathsf T}J\xi}\,dz.
\end{equation}
We obtain
\begin{equation}
\hat A_{\boldsymbol\Delta}
=\frac{1}{2\pi\hbar}\int_{\mathbb R^2}
\widetilde A_{\boldsymbol\Delta}(\xi)\;
e^{-\frac14\,\xi^{\mathsf T}\Sigma^{-2}\xi}\;
D(\xi)\,d\xi.
\label{eq:ADelta_Weyl_rewrite}
\end{equation}

Since $A_{\boldsymbol\Delta}=G_{\boldsymbol\Delta}*A$, the convolution theorem gives
$\widetilde A_{\boldsymbol\Delta}=(2\pi\hbar)\,\widetilde G_{\boldsymbol\Delta}\,\widetilde A$.
Absorbing ${2\pi \hbar\,}\widetilde G_{\boldsymbol\Delta}$ and the Gaussian factor in Eq.~\eqref{eq:ADelta_Weyl_rewrite} into a single damping function yields
\begin{equation}
\hat A_{\boldsymbol\Delta}
=\frac{1}{2\pi\hbar}\int_{\mathbb R^2}\widetilde A(\xi)\,
\phi_{\boldsymbol\Delta}(\xi)\,D(\xi)\,d\xi,
\label{eq:Achar_rewrite}
\end{equation}
where
\begin{equation}
\phi_{\boldsymbol \Delta}(\xi):=
e^{-\xi^{\mathsf T}\Gamma_\Delta\xi},
\qquad
\Gamma_{\boldsymbol \Delta}
:=\mathrm{diag}\!\left(
\frac{1}{4\sigma_x^2}+\frac{\Delta_p^2\sigma_p^2}{4\hbar^2},
\;
\frac{1}{4\sigma_p^2}+\frac{\Delta_x^2\sigma_x^2}{4\hbar^2}
\right).
\end{equation}
Using Eq.~\eqref{eq:Achar_rewrite} for both $\hat A_{\boldsymbol\Delta}$ and $\hat B_{\boldsymbol\Delta}$, we obtain
\begin{equation}
[\hat A_{\boldsymbol\Delta},\hat B_{\boldsymbol\Delta}]
=\frac{1}{(2\pi\hbar)^2}\iint_{\mathbb{R}^2\times\mathbb{R}^2}
\widetilde{A}(\xi_1)\widetilde{B}(\xi_2)\,
\phi_{\boldsymbol\Delta}(\xi_1)\phi_{\boldsymbol\Delta}(\xi_2)\,
\big[D(\xi_1),D(\xi_2)\big]\,
d\xi_1\,d\xi_2.
\label{eq:comm_start_rewrite}
\end{equation}
The Weyl relations imply
\begin{equation}
D(\xi_1)D(\xi_2)
=e^{\frac{i}{2\hbar}(\xi_1^{\mathsf T}J\xi_2)}\,D(\xi_1+\xi_2),
\end{equation}
hence
\begin{equation}
[D(\xi_1),D(\xi_2)]
=2i\sin\!\Big(\frac{\xi_1^{\mathsf T}J\xi_2}{2\hbar}\Big)\,D(\xi_1+\xi_2).
\end{equation}
Since $D(\xi)$ is unitary, $\|D(\xi)\|=1$, and therefore
\begin{equation}
\big\|[D(\xi_1),D(\xi_2)]\big\|
=2\Big|\sin\!\Big(\frac{\xi_1^{\mathsf T}J\xi_2}{2\hbar}\Big)\Big|
\le \frac{|\xi_1^{\mathsf T}J\xi_2|}{\hbar}.
\label{eq:Dcomm_bound_rewrite}
\end{equation}
Taking operator norms in Eq.~\eqref{eq:comm_start_rewrite} and using Eq.~\eqref{eq:Dcomm_bound_rewrite} gives
\begin{equation}
\big\|[\hat A_{\boldsymbol\Delta},\hat B_{\boldsymbol\Delta}]\big\|
\le
\frac{1}{(2\pi\hbar)^2\,\hbar}\iint_{\mathbb{R}^2\times\mathbb{R}^2}
|\widetilde{A}(\xi_1)|\,|\widetilde{B}(\xi_2)|\,
\phi_{\boldsymbol\Delta}(\xi_1)\phi_{\boldsymbol\Delta}(\xi_2)\,
|\xi_1^{\mathrm T}J\xi_2|\,
d\xi_1\,d\xi_2 .
\label{eq:comm_bound1_rewrite}
\end{equation}

Next, by the triangle inequality applied to the inverse transform defining $\widetilde A$,
\begin{equation}
|\widetilde A(\xi)|\le \frac{1}{2\pi\hbar}\int_{\mathbb R^2}|A(z)|\,dz
=: \|A\|_{L^1_\hbar},
\qquad
|\widetilde B(\xi)|\le \|B\|_{L^1_\hbar}.
\end{equation}
Thus the $\xi$-dependent part is separated, and the remaining Gaussian integral can be evaluated explicitly:
\begin{equation}
\iint_{\mathbb{R}^2\times\mathbb{R}^2}
\phi_{\boldsymbol\Delta}(\xi_1)\phi_{\boldsymbol\Delta}(\xi_2)\,
|\xi_1^{\mathrm T}J\xi_2|\,
d\xi_1\,d\xi_2
=
\frac{4\pi^2\hbar^3}{
\left(1+\frac{\Delta_x^2}{4}\right)^{3/2}
\left(1+\frac{\Delta_p^2}{4}\right)^{3/2}}.
\end{equation}
Substituting into Eq.~\eqref{eq:comm_bound1_rewrite} yields
\begin{equation}
\big\|[\hat A_{\boldsymbol\Delta},\hat B_{\boldsymbol\Delta}]\big\|
\le
\frac{\|A\|_{L^1_\hbar}\,\|B\|_{L^1_\hbar}}{
\left(1+\frac{\Delta_x^2}{4}\right)^{3/2}
\left(1+\frac{\Delta_p^2}{4}\right)^{3/2}},
\end{equation}
which is the desired bound.
\subsection{Statement 3}\label{app:statement3}
Consider the differential equation
\begin{equation}
    \partial_t p_{\boldsymbol \Delta} - \partial_t p_{\boldsymbol \Delta}^\text{cl}
    = L_{H_{\boldsymbol \Delta}}(p_{\boldsymbol \Delta} - p_{\boldsymbol \Delta}^\text{cl})
    + (\mathcal D_{\boldsymbol\Delta}+\mathcal{R}_{\boldsymbol\Delta})p_{\boldsymbol\Delta}.
\end{equation}
We can solve it with Duhamel's formula, obtaining
\begin{equation}\label{eq:Duhamel_Liouville}
p_{\boldsymbol\Delta}(t)-p_{\boldsymbol\Delta}^\text{cl}(t)
=
\int_0^t \Phi^{\,t-s}\,
\big[(\mathcal D_{\boldsymbol\Delta}+\mathcal{R}_{\boldsymbol\Delta})p_{\boldsymbol\Delta}(s)\big]\,ds,
\end{equation}
where $\Phi^{t}$ denotes the propagator generated by $L_{H_{\boldsymbol\Delta}}$.

Since $\Phi^{t} f=f\circ\varphi^{-t}$ for the Hamiltonian flow $\varphi^{t}$ of $H_{\boldsymbol\Delta}$,
it preserves $L^\infty$ norms up to transport of the domain:
$\|\Phi^{t} f\|_{L^\infty(K)}=\|f\|_{L^\infty(\varphi^{-t}(K))}$.
Therefore, for any bounded $K\subset\mathbb R^2$ and $K_t:=\varphi^{-t}(K)$,
\begin{equation}
\|p_{\boldsymbol\Delta}(t)-p_{\boldsymbol\Delta}^\text{cl}(t)\|_{L^\infty(K)}
\le
\int_0^t
\|(\mathcal D_{\boldsymbol\Delta}+\mathcal{R}_{\boldsymbol\Delta})p_{\boldsymbol\Delta}(s)\|_{L^\infty(K_{t-s})}\,ds,
\end{equation}
which can be separated in the three terms concluding the proof.

\subsection{Statement 4}\label{app:statement4}
Starting from the definition~\eqref{eq:RDelta_def_min} with $f=p_{\boldsymbol\Delta}$,
\begin{equation}
\mathcal{R}_{\boldsymbol\Delta} p_{\boldsymbol\Delta}
:=S_{\boldsymbol\Delta}\!\left(\frac{2}{\hbar}\,H\,\sin\!\Big(\frac{\hbar}{2}\Lambda\Big)\,S_{\boldsymbol\Delta}^{-1}p_{\boldsymbol\Delta}
-\{H,S_{\boldsymbol\Delta}^{-1}p_{\boldsymbol\Delta}\}_{\mathrm{PB}}\right),
\qquad
\Lambda:=\overleftarrow{\partial}_x\,\overrightarrow{\partial}_p-\overleftarrow{\partial}_p\,\overrightarrow{\partial}_x ,
\end{equation}
expand the sine in a power series,
\begin{equation}
\frac{2}{\hbar}\sin\!\Big(\frac{\hbar}{2}\Lambda\Big)
=\sum_{n=0}^{\infty}\frac{(-1)^n}{(2n+1)!}\Big(\frac{\hbar}{2}\Big)^{2n}\Lambda^{2n+1},
\end{equation}
and note that the Poisson bracket subtracts the $n=0$ term. For $N\in\mathbb N$, this yields the partial remainder
\begin{equation}
\mathcal{R}_{\boldsymbol\Delta}^{(N)}p_{\boldsymbol\Delta}
=
\sum_{n=1}^{N}
\frac{(-1)^n}{(2n+1)!}\Big(\frac{\hbar}{2}\Big)^{2n}\,
S_{\boldsymbol\Delta}\!\Big[(\partial_x^{2n+1}V)\,\partial_p^{2n+1}W\Big].
\end{equation}

Fix a region $K\subset\mathbb R^2$. Recall from Eq.~\eqref{eq:pDelta_smoothing} that
$S_{\boldsymbol\Delta}g = G_{\boldsymbol\Delta}*g$.
Assume $W$ is sufficiently regular/decaying so that the following identity is legitimate
(e.g.\ $W\in\mathcal S(\mathbb R^2)$). Then, for each $n$,
\begin{equation}
S_{\boldsymbol\Delta}\!\Big[(\partial_x^{2n+1}V)\,\partial_p^{2n+1}W\Big]
=
(\partial_p^{2n+1}G_{\boldsymbol\Delta})*\Big[(\partial_x^{2n+1}V)\,W\Big],
\end{equation}
since differentiation commutes with convolution in the distributional sense.

Using $\|h\|_{L^\infty(K)}\le \|h\|_{L^\infty(\mathbb R^2)}$ and Young's inequality
$\|k*f\|_{L^\infty}\le \|k\|_{L^\infty}\|f\|_{L^1}$, we obtain
\begin{equation}
\Big\|S_{\boldsymbol\Delta}\!\big[(\partial_x^{2n+1}V)\,\partial_p^{2n+1}W\big]\Big\|_{L^\infty(K)}
\le
\|\partial_p^{2n+1}G_{\boldsymbol\Delta}\|_{L^\infty}\;
\|(\partial_x^{2n+1}V)\,W\|_{L^1(\mathbb R^2)}.
\end{equation}
Hence, by the triangle inequality,
\begin{equation}
\|\mathcal{R}_{\boldsymbol\Delta}^{(N)}p_{\boldsymbol\Delta}\|_{L^\infty(K)}
\le
\sum_{n=1}^{N}\frac{1}{(2n+1)!}\Big(\frac{\hbar}{2}\Big)^{2n}\,
\|\partial_p^{2n+1}G_{\boldsymbol\Delta}\|_{L^\infty}\;
\|(\partial_x^{2n+1}V)\,W\|_{L^1}.
\label{eq:RDelta_bound_shift_N}
\end{equation}

For the Gaussian kernel $G_{\boldsymbol\Delta}$, derivatives in $p$ can be expressed in terms of
(probabilists') Hermite polynomials $\mathrm{He}_k$, defined by
\begin{equation}
\mathrm{He}_k(u):=(-1)^k e^{u^2/2}\frac{d^k}{du^k}e^{-u^2/2}.
\end{equation}
Writing the $p$-dependence of $G_{\boldsymbol\Delta}$ as
$
\exp\!\big(-p^2/(2\ell_p^2)\big)
$
with $\ell_p^2=\sigma_p^2(\Delta_p^2+1)$, one has
\begin{equation}
\partial_p^{k}G_{\boldsymbol\Delta}(x,p)
=
\frac{(-1)^k}{\ell_p^{k}}\,
\mathrm{He}_k\!\Big(\frac{p}{\ell_p}\Big)\,G_{\boldsymbol\Delta}(x,p),
\qquad k\ge 0,
\end{equation}
and therefore
\begin{equation}
\|\partial_p^{k}G_{\boldsymbol\Delta}\|_{L^\infty}
\le
\frac{\kappa_k}{\ell_p^{k}}\;\|G_{\boldsymbol\Delta}\|_{L^\infty},
\qquad
\kappa_k:=\sup_{u\in\mathbb R}\big|\mathrm{He}_k(u)\big|e^{-u^2/2},
\qquad
\|G_{\boldsymbol\Delta}\|_{L^\infty}=\frac{1}{2\pi\,\ell_x\ell_p}.
\label{eq:Gaussian_derivative_bound_kappa_He}
\end{equation}
Combining Eqs.~\eqref{eq:RDelta_bound_shift_N} and~\eqref{eq:Gaussian_derivative_bound_kappa_He} yields
\begin{equation}
\|\mathcal{R}_{\boldsymbol\Delta}^{(N)}p_{\boldsymbol\Delta}\|_{L^\infty(K)}
\le
\frac{1}{2\pi\,\ell_x\ell_p}
\sum_{n=1}^{N}\frac{\kappa_{2n+1}}{(2n+1)!}\Big(\frac{\hbar}{2}\Big)^{2n}\,
\frac{1}{\ell_p^{2n+1}}\;
\|(\partial_x^{2n+1}V)\,W\|_{L^1(\mathbb R^2)}.
\label{eq:RDelta_bound_shift_series_He}
\end{equation}

Assume in addition that the series obtained by letting $N\to\infty$ on the right-hand side of Eq.~\eqref{eq:RDelta_bound_shift_series_He}
is convergent. Then $\{\mathcal{R}_{\boldsymbol\Delta}^{(N)}p_{\boldsymbol\Delta}\}_{N\in\mathbb N}$ is Cauchy in $L^\infty(K)$ and defines
\begin{equation}
\mathcal{R}_{\boldsymbol\Delta}p_{\boldsymbol\Delta}:=\lim_{N\to\infty}\mathcal{R}_{\boldsymbol\Delta}^{(N)}p_{\boldsymbol\Delta}\quad\text{in }L^\infty(K),
\end{equation}
and Eq.~\eqref{eq:RDelta_bound_shift_series_He} holds with $N=\infty$, implying
\begin{equation}
\|\mathcal{R}_{\boldsymbol\Delta}p_{\boldsymbol\Delta}\|_{L^\infty(K)}\longrightarrow 0
\qquad\text{as}\qquad
\,\ell_x\ell_p\longrightarrow\infty.
\end{equation}
\medskip
The assumptions used above are that the identity
$
G_{\boldsymbol\Delta}*\big[(\partial_x^{2n+1}V)\,\partial_p^{2n+1}W\big]
=(\partial_p^{2n+1}G_{\boldsymbol\Delta})*\big[(\partial_x^{2n+1}V)\,W\big]
$
is legitimate (e.g.\ $W$ sufficiently regular/decaying so that integration by parts in $p$ produces no boundary terms),
that $(\partial_x^{2n+1}V)\,W\in L^1(\mathbb R^2)$ for the relevant $n$, and that the series on the right-hand side of
Eq.~\eqref{eq:RDelta_bound_shift_series_He} converges when $N\to\infty$. These requirements are met in many standard situations,
for instance:
\begin{itemize}
\item \emph{Periodic/bounded potentials and Schwartz $W$.}
If $V\in C_b^\infty(\mathbb R)$ (all derivatives $\partial_x^k V$ are bounded; e.g.\ $V(x)=\sin x$, $\cos x$, and other smooth
periodic potentials) and $W\in\mathcal S(\mathbb R^2)$, then $(\partial_x^{2n+1}V)\,W\in L^1(\mathbb R^2)$ for all $n$, the
convolution-by-parts identity holds, and the bound series is well behaved for fixed $\boldsymbol\Delta$ and in particular for large
$\Delta_p$.

\item \emph{Polynomial-growth potentials and rapidly decaying $W$.}
If $\partial_x^kV$ grows at most polynomially in $x$ (for all $k$) and $W$ decays sufficiently fast in $x$ (e.g.\ Schwartz in $x$),
then $(\partial_x^{2n+1}V)\,W\in L^1(\mathbb R^2)$ for all $n$. If moreover $W$ is Schwartz in $p$, the integration-by-parts step
is justified.

\item \emph{Polynomial potentials.}
If $V$ is a polynomial, then $\partial_x^{2n+1}V\equiv 0$ for $n$ large enough, so the expansion truncates and the $N=\infty$ issue
does not arise.
\end{itemize}

\subsection{Statement 5}\label{app:statement5}
Starting from the definition~\eqref{eq:DDelta_def_min} with $f=p_{\boldsymbol\Delta}$ and writing
$W=S_{\boldsymbol\Delta}^{-1}p_{\boldsymbol\Delta}$,
\begin{equation}
\mathcal D_{\boldsymbol\Delta}p_{\boldsymbol\Delta}
=
S_{\boldsymbol\Delta}\!\left(\{H,W\}_{\mathrm{PB}}\right)
-\{S_{\boldsymbol\Delta}H,p_{\boldsymbol\Delta}\}_{\mathrm{PB}}.
\end{equation}
Fix a bounded region $K\subset\mathbb R^2$ and take $H(x,p)=\frac{p^2}{2m}+V(x)$. Assume $W$ is sufficiently regular/decaying so that the
integrations by parts below are legitimate and $\|W\|_{L^1(\mathbb R^2)},\ \|V'W\|_{L^1(\mathbb R^2)}<\infty$. For this Hamiltonian,
\begin{equation}
\{H,W\}_{\mathrm{PB}}=V'(x)\,\partial_p W-\frac{p}{m}\,\partial_x W,
\qquad
\{S_{\boldsymbol\Delta}H,p_{\boldsymbol\Delta}\}_{\mathrm{PB}}
=(S_{\boldsymbol\Delta}V')\,\partial_p p_{\boldsymbol\Delta}-\frac{p}{m}\,\partial_x p_{\boldsymbol\Delta},
\end{equation}
hence
\begin{equation}\label{eq:D_split_identity}
\mathcal D_{\boldsymbol\Delta}p_{\boldsymbol\Delta}
=
\Big(S_{\boldsymbol\Delta}(V'\partial_p W)-(S_{\boldsymbol\Delta}V')\,\partial_p p_{\boldsymbol\Delta}\Big)
-\frac{1}{m}\Big(S_{\boldsymbol\Delta}(p\,\partial_x W)-p\,\partial_x p_{\boldsymbol\Delta}\Big).
\end{equation}
Let us write $S_{\boldsymbol\Delta}=S_xS_p$ with $
S_x:=e^{\frac{\ell_x^2}{2}\partial_x^2},\, S_p:=e^{\frac{\ell_p^2}{2}\partial_p^2}.
$ Using $[\partial_p^2,p]=2\partial_p$, one has
$e^{\frac{\ell_p^2}{2}\partial_p^2}\,p=(p+\ell_p^2\partial_p)e^{\frac{\ell_p^2}{2}\partial_p^2}$ and therefore
$S_{\boldsymbol\Delta}(p\,g)=p\,S_{\boldsymbol\Delta}g+\ell_p^2\partial_p(S_{\boldsymbol\Delta}g)$, which with $g=\partial_x W$ gives
\begin{equation}\label{eq:D_kinetic_exact}
S_{\boldsymbol\Delta}(p\,\partial_x W)-p\,\partial_x p_{\boldsymbol\Delta}
=\ell_p^2\,\partial_{xp}p_{\boldsymbol\Delta}.
\end{equation}

Since $p_{\boldsymbol\Delta}=G_{\boldsymbol\Delta}*W$, by Young's inequality
\begin{equation}
\|\partial_{xp}p_{\boldsymbol\Delta}\|_{L^\infty(K)}
\le
\|\partial_{xp}G_{\boldsymbol\Delta}\|_{L^\infty(\mathbb R^2)}\,\|W\|_{L^1(\mathbb R^2)}.
\end{equation}
Furthermore $\|\partial_{xp}G_{\boldsymbol\Delta}\|_{L^\infty(\mathbb R^2)}
=
\frac{1}{2\pi e}\,\frac{1}{\ell_x^2\,\ell_p^2},
$ and thus
\begin{equation}\label{eq:D_kinetic_bound_final}
\left\|\frac{1}{m}\Big(S_{\boldsymbol\Delta}(p\,\partial_x W)-p\,\partial_x p_{\boldsymbol\Delta}\Big)\right\|_{L^\infty(K)}
\le
\frac{1}{2\pi e\,m}\,\frac{1}{\ell_x^2}\,\|W\|_{L^1(\mathbb R^2)}.
\end{equation}
For the potential term, 
since $V'(x)$ depends only on $x$, it commutes with $S_p$, and $S_p$ commutes with $\partial_p$. Setting $q:=S_pW$ (so that
$p_{\boldsymbol\Delta}=S_xq$) we obtain the exact identity
\begin{align}\label{eq:D_pot_reduce_nom}
S_{\boldsymbol\Delta}(V'\partial_pW)-(S_{\boldsymbol\Delta}V')\,\partial_pp_{\boldsymbol\Delta}
&=
S_x\!\big(V'\,\partial_p q\big)-(S_xV')\,\partial_p(S_xq)\nonumber\\
&=
S_x\!\big(V'\,S_x^{-1}(\partial_p p_{\boldsymbol\Delta})\big)-(S_xV')\,\partial_p p_{\boldsymbol\Delta}.
\end{align}

Using $[\partial_x^2,g]=2g'\partial_x+g''$ (as operators acting on test functions) and the Hadamard lemma, one has the (formal)
conjugation expansion
\begin{equation}
S_x\!\big(g\,S_x^{-1}u\big)
=
\sum_{n\ge0}\frac{1}{n!}\,\big(\partial_x^n(S_xg)\big)\,(\ell_x^2\partial_x)^n u,
\end{equation}
so that subtracting the $n=0$ term yields the defect formula
\begin{equation}
S_x\!\big(g\,S_x^{-1}u\big)-(S_xg)\,u
=
\sum_{n\ge1}\frac{(\ell_x^2)^n}{n!}\,\big(\partial_x^n(S_xg)\big)\,\partial_x^n u.
\end{equation}
Applying this with $g=V'$ and using $\partial_x^n(S_xV')=S_x(\partial_x^nV')=S_xV^{(n+1)}$, and then setting
$u=\partial_p p_{\boldsymbol\Delta}$, we obtain
\begin{equation}\label{eq:D_potential_exact_series_nom}
S_{\boldsymbol\Delta}(V'\partial_pW)-(S_{\boldsymbol\Delta}V')\,\partial_pp_{\boldsymbol\Delta}
=
\sum_{n\ge1}\frac{(\ell_x^2)^n}{n!}\,\big(S_xV^{(n+1)}\big)\,\partial_x^n\partial_p p_{\boldsymbol\Delta}.
\end{equation}
(In particular, if $V$ is a polynomial then the sum truncates at finite $n$.)

Since $p_{\boldsymbol\Delta}=G_{\boldsymbol\Delta}*W$, we have $
\partial_x^n\partial_p p_{\boldsymbol\Delta}
=
(\partial_x^n\partial_pG_{\boldsymbol\Delta})*W,
$
and by Young's inequality and $\|h\|_{L^\infty(K)}\le \|h\|_{L^\infty(\mathbb R^2)}$,
\begin{equation}\label{eq:D_potential_derivative_young_nom}
\|\partial_x^n\partial_p p_{\boldsymbol\Delta}\|_{L^\infty(K)}
\le
\|\partial_x^n\partial_pG_{\boldsymbol\Delta}\|_{L^\infty}\;\|W\|_{L^1(\mathbb R^2)}.
\end{equation}
To bound the Gaussian derivatives, note that for $k\ge 0$,
\begin{equation}
\partial_x^{k}G_{\boldsymbol\Delta}(x,p)
=
\frac{(-1)^k}{\ell_x^{k}}\,
\mathrm{He}_k\!\Big(\frac{x}{\ell_x}\Big)\,G_{\boldsymbol\Delta}(x,p),
\qquad
\partial_pG_{\boldsymbol\Delta}(x,p)
=
-\frac{1}{\ell_p}\,
\mathrm{He}_1\!\Big(\frac{p}{\ell_p}\Big)\,G_{\boldsymbol\Delta}(x,p),
\end{equation}
and therefore
\begin{equation}
\partial_x^n\partial_pG_{\boldsymbol\Delta}(x,p)
=
\frac{(-1)^{n+1}}{\ell_x^{n}\ell_p}\,
\mathrm{He}_n\!\Big(\frac{x}{\ell_x}\Big)\,
\mathrm{He}_1\!\Big(\frac{p}{\ell_p}\Big)\,G_{\boldsymbol\Delta}(x,p).
\end{equation}
Using the definition $\kappa_k:=\sup_{u\in\mathbb R}|\mathrm{He}_k(u)|e^{-u^2/2}$ from Eq.~\eqref{eq:Gaussian_derivative_bound_kappa_He} and $\|G_{\boldsymbol\Delta}\|_{L^\infty}=(2\pi\ell_x\ell_p)^{-1}$, we obtain
\begin{equation}\label{eq:D_kernel_sup_general_nom}
\|\partial_x^n\partial_pG_{\boldsymbol\Delta}\|_{L^\infty}
\le
\frac{\kappa_n\,\kappa_1}{\ell_x^{n}\ell_p}\;\|G_{\boldsymbol\Delta}\|_{L^\infty}=
\frac{1}{2\pi\,\ell_x\ell_p}\,\frac{\kappa_n\,\kappa_1}{\ell_x^{n}\ell_p}.
\end{equation}
Combining Eqs.~\eqref{eq:D_potential_exact_series_nom}, \eqref{eq:D_potential_derivative_young_nom} and
\eqref{eq:D_kernel_sup_general_nom}, and bounding term-by-term yields
\begin{align}
\|S_{\boldsymbol\Delta}(V'\partial_pW)-(S_{\boldsymbol\Delta}V')\,\partial_pp_{\boldsymbol\Delta}\|_{L^\infty(K)}
&\le
\sum_{n\ge1}\frac{(\ell_x^2)^n}{n!}\,
\|S_xV^{(n+1)}\|_{L^\infty(K)}\,
\|\partial_x^n\partial_p p_{\boldsymbol\Delta}\|_{L^\infty(K)}\nonumber\\
&\le
\frac{\kappa_1}{2\pi\,\ell_x\ell_p^2}\,\|W\|_{L^1(\mathbb R^2)}
\sum_{n\ge1}\frac{\kappa_n}{n!}\,\ell_x^{\,n}\,\|S_xV^{(n+1)}\|_{L^\infty(K)}.
\label{eq:D_potential_bound_series_nom}
\end{align}
In particular, the leading ($n=1$) shear term satisfies (using $\kappa_1=e^{-1/2}$ and $\|\partial_{xp}G_{\boldsymbol\Delta}\|_{L^\infty}
=(2\pi e)^{-1}(\ell_x^2\ell_p^2)^{-1}$)
\begin{equation}\label{eq:D_potential_bound_leading_nom}
\|(\ell_x^2)(S_xV'')\,\partial_{xp}p_{\boldsymbol\Delta}\|_{L^\infty(K)}
\le
\frac{1}{2\pi e}\,\frac{1}{\ell_p^2}\,\|S_xV''\|_{L^\infty(K)}\,\|W\|_{L^1(\mathbb R^2)}.
\end{equation}

Finally, combining Eqs.~\eqref{eq:D_split_identity}, \eqref{eq:D_kinetic_bound_final}, and \eqref{eq:D_potential_bound_series_nom} yields
\begin{equation}\label{eq:DDelta_bound_shear_series_nom}
\|\mathcal D_{\boldsymbol\Delta}p_{\boldsymbol\Delta}\|_{L^\infty(K)}
\le
\frac{\kappa_1}{2\pi\,\ell_x\ell_p^2}\,\|W\|_{L^1(\mathbb R^2)}
\sum_{n\ge1}\frac{\kappa_n}{n!}\,\ell_x^{\,n}\,\|S_xV^{(n+1)}\|_{L^\infty(K)}
+
\frac{1}{2\pi e\,m}\,\frac{1}{\ell_x^2}\,\|W\|_{L^1(\mathbb R^2)}.
\end{equation}
\subsubsection{Absolute convergence of the Hermite series in Statement~\ref{st:D_bound}}
\label{app:absolute_convergence}

Set
\begin{equation}
\epsilon:=\frac{\ell_x}{L_V(K)}\ge 0.
\end{equation}
We show that
\begin{equation}
\sum_{n\ge1}\frac{\kappa_n}{n!}\epsilon^{\,n-1},
\qquad
\kappa_n:=\sup_{u\in\mathbb R} |\mathrm{He}_n(u)|e^{-u^2/2},
\end{equation}
converges absolutely for every finite $\epsilon$.

Using the generating function
\begin{equation}
e^{ut-t^2/2}=\sum_{n=0}^\infty \frac{\mathrm{He}_n(u)}{n!}t^n,
\end{equation}
we get
\begin{equation}
e^{-u^2/2}e^{ut-t^2/2}=e^{-(u-t)^2/2}
=\sum_{n=0}^\infty \frac{\mathrm{He}_n(u)e^{-u^2/2}}{n!}t^n.
\end{equation}
Thus, for fixed $u\in\mathbb R$, the coefficient
\begin{equation}
\frac{\mathrm{He}_n(u)e^{-u^2/2}}{n!}
\end{equation}
is the $n$th Taylor coefficient of the entire function
\begin{equation}
f_u(t):=e^{-(u-t)^2/2}.
\end{equation}
By Cauchy's estimate, for every $R>0$,
\begin{equation}
\frac{|\mathrm{He}_n(u)|e^{-u^2/2}}{n!}
\le
R^{-n}\sup_{|t|=R}|f_u(t)|.
\end{equation}
Now, if $t=a+ib$ and $u\in\mathbb R$, then
\begin{equation}
|f_u(t)|
=
\exp\!\left(-\frac12\Re((u-t)^2)\right)
=
\exp\!\left(-\frac{(u-a)^2}{2}+\frac{b^2}{2}\right)
\le e^{R^2/2}
\qquad (|t|=R).
\end{equation}
Hence
\begin{equation}
\frac{\kappa_n}{n!}\le e^{R^2/2}R^{-n}
\qquad \forall R>0.
\end{equation}
Choosing $R=\sqrt n$ gives
\begin{equation}
\frac{\kappa_n}{n!}\le \left(\frac{e}{n}\right)^{n/2}.
\end{equation}
Therefore
\begin{equation}
\left(\frac{\kappa_n}{n!}\epsilon^{\,n-1}\right)^{1/n}
\le
\epsilon^{(n-1)/n}\left(\frac{e}{n}\right)^{1/2}\to 0,
\end{equation}
so the series converges absolutely for every finite $\epsilon$.

In particular,
\begin{equation}
\sum_{n\ge1}\frac{\kappa_n}{n!}\epsilon^{\,n-1}
=
\kappa_1+O(\epsilon),
\qquad \epsilon\to0.
\end{equation}

\bibliographystyle{alpha}
\bibliography{main}

\end{document}